\documentclass[11pt,a4paper]{article}
\usepackage{amsfonts,amsmath,amssymb,mathrsfs}
\usepackage{color,epsfig,graphicx,graphics}
\usepackage{hyperref}
\usepackage{braket}

\newtheorem{theorem}{Theorem}[section]

\newtheorem{corollary}{Corollary}[section]

\newtheorem{lemma}{Lemma}[section]

\newtheorem{remark}{Remark}[section]
\numberwithin{equation}{section}

\newcommand{\beq}{\begin{equation}}
\newcommand{\eeq}{\end{equation}}
\newcommand{\beqn}{\begin{eqnarray}}
\newcommand{\eeqn}{\end{eqnarray}}
\newcommand{\bgn}{\begin{align}}
\newcommand{\egn}{\end{align}}
\newcommand{\bmrk}{\begin{remark}}
\newcommand{\emrk}{\end{remark}}
\newcommand{\beqm}{\begin{equation*}}
\newcommand{\eeqm}{\end{equation*}}
\newcommand{\beqnm}{\begin{eqnarray*}}
\newcommand{\eeqnm}{\end{eqnarray*}}
\newcommand{\bgnm}{\begin{align*}}
\newcommand{\egnm}{\end{align*}}

\begin{document}

\title{On the response of  a two-level system to two-photon inputs}

\author{Zhiyuan Dong\thanks{Shenzhen Graduate School, Harbin Institute of Technology, Shenzhen 518055, China 
  (zhiyuan.dong@connect.polyu.hk).}
\and Guofeng Zhang\thanks{Corresponding author. Department of Applied Mathematics, Hong Kong Polytechnic University, Hong Kong, China
  (guofeng.zhang@polyu.edu.hk, \url{https://www.polyu.edu.hk/ama/profile/gfzhang/}).}
\and Nina H. Amini\thanks{Laboratoire des signaux et syst\`{e}mes(L2S), CNRS-CentraleSup\'{e}lec-Universit\'{e} Paris-Sud, Universit\'{e} Paris-Saclay, 3, rue Joliot Curie, 91190 Gif-sur-Yvette, France
  (nina.amini@l2s.centralesupelec.fr).}
}
\maketitle

\begin{abstract}
The purpose of this paper is to  study the interaction  between a two-level system (qubit) and two continuous-mode photons. Two scenarios are investigated: Case 1, how a two-level system changes the pulse shapes of two  input photons propagating in a single input channel; and Case 2, how a two-level system responds to two counter-propagating photons, one in each input channel. By means of a transfer function approach, the steady-state output field states for both cases are derived analytically in both the time and frequency domains. For Case 1, two examples are presented. In Example 1 a two-photon input state of Gaussian pulse shape is used to excite a two-level atom. The joint probability distribution in the time domain and the joint spectra of the output two-photon state are plotted. The simulation demonstrates that in the time domain the atom tends to stretch out the two photons. Moreover, the prominent difference between the  joint probability distribution of the output two-photon state  and that of the input two-photon state  occurs exactly under the setting when the two-level atom is most efficiently excited. In Example 2,  a two-photon input state of rising exponential pulse shape is used to excite a two-level atom. Strong anti-correlation of the  output two-photon state is observed, which is absent in Example 1 for the Gaussian pulse shape.  Such difference indicates that different pulse shapes give rise to drastically different frequency entanglement of the output two-photon state.  Example 3 is used to illustrate Case 2, where two counter-propagating single photons of rising exponential pulse shapes are input to a two-level atom. The \textit{frequency-dependent} Hong-Ou-Mandel (HOM) interference phenomenon  is observed. Moreover, when the two output photons are in the same channel, they are anti-correlated.  The simulation results base on the analytic forms of output two-photon states are consistent with those based on quantum master or filter equations \cite{SONG:13MULTI}, \cite{DZA2019}. Similar physical phenomena have been observed in physical settings such as cavity opto-mechanical systems and Keer nonlinear cavities.  
\end{abstract}

\textbf{keywords.}
Quantum control, two-level system, input-output formalism, transfer function

%%%%%%%%%%%%%%%%%%%%%%%%
%%%%%%%%%%%%%%%%%%%%%%%%
%%%%%%%%%%%%%%%%%%%%%%%%
\section{Introduction}\label{sec:intro}

Strong coupling of a two-level quantum system to a quantized radiation field can give rise to rich and interesting physical phenomena. Strong coupling can be achieved in various physical setups, for example, by putting an atom in a cavity (cavity QED), by embedding a two-level emitter in a nanophotonic waveguide (waveguide QED), or by coupling a superconducting qubit to a transmission line resonator (circuit QED). In the strong coupling regime, the pulse shape of a photon, which specifies the energy distribution of the photon around the carrier frequency, has a remarkable influence on the interaction between the photon and the two-level system. For example, a two-level atom, initially in the ground state, can be fully excited by a single photon of rising exponential pulse shape provided that the photon's full width at half maximum (FWHM) $\gamma$ is equal to the decay rate $\kappa$ of the atom \cite{stobinska2009perfect}, \cite{WMSS:11}, \cite{PZJ:15}. In contrast, if the incident photon is of Gaussian pulse shape with frequency bandwidth $\Omega$, the maximal excitation probability is $0.8$, which is attained at $\Omega=1.46\kappa$; see, e.g., \cite{stobinska2009perfect}, \cite{RSF10}, \cite[Fig. 1]{WMSS:11}, \cite[Fig. 8]{GJNC:12}, \cite[Fig. 2]{baragiola:2012}. Recently, an analytical expression of the output single photon state for a two-level atom driven by a single photon has been derived in \cite{PZJ:15}. Assume that the Gaussian pulse shape $\xi(t)$ of the input photon has the photon peak arrival time $\tau=3$ and frequency bandwidth $\Omega=1.46\kappa$. Denote the pulse shape of the output single photon by $\eta(t)$. Then it can be found that $\int_{-\infty}^{4}\left(|\xi(r)|^2-|\eta(r)|^2\right)dr=0.8$. Interestingly, the excitation probability of the atom achieves its maximum $0.8$ also at time $t=4$. For more discussions of the interaction between a single photon and a two-level quantum system in various physical setups, interested reader may refer to \cite{SF:05,WMSS:11,Rep:12,GJN:13,ZGPG18} and references therein.

The dynamics of a two-level system driven by a two-photon wavepacket are much more complicated. When a two-level system is driven by two photons in a single input channel, in other words, the photons can only propagate along one direction, e.g., in a chiral waveguide \cite{SF:05,Shen:07,FKS:10}, the scattering matrix ($S$-matrix) has been derived explicitly in \cite{FKS:10,Xu:13}. In \cite{SONG:13MULTI}, quantum filters for a Markovian quantum system driven by an arbitrary number of photons in a single channel have been derived. As demonstration, the atomic excitation of a two-level atom driven by a two-photon state of Gaussian pulse shape has been studied. Numerical simulations show that the maximal excitation probability is $0.8796$ when the frequency bandwidths $\Omega_1=\Omega_2=2*1.46\kappa$, see \cite[Fig. 1]{SONG:13MULTI}. In \cite{Nysteen:14}, the scattering of two photons on a quantum two-level emitter embedded in a one-dimensional waveguide is considered, where it is found that photon transport depends on the excitation of the emitter. Moreover, the authors of \cite{Nysteen:14} also studied the correlation and entanglement between the two output  photons induced by a two-level emitter which is driven by two counter-propagating input photon pulses. The effect of the pulse shapes of the two counter-propagating input photons on the induced correlations of the two output photons is studied in \cite{Nysteen15}; to be specific, the output two-photon state is derived, based on which the output intensity spectra are investigated when the input photons are of Gaussian pulse shapes with various spectral widths. In \cite{Roulet16}, time and frequency correlations between the two output photons are investigated. Moreover, the relationship between induced photon-photon correlations and the atomic excitation efficiency is analyzed. When a two-level system is driven by two counter-propagating indistinguishable single photons, it is shown in \cite{DZA2019} that the maximal excitation probability attains at $\gamma=5\kappa$ for rising exponential pulse shapes, and $\Omega=2*1.46\kappa$ for Gaussian pulse shapes. Recently, the dynamics of two two-level systems (qubits) driven by two counter-propagating input photons is studied in \cite{ZP:19}. Based on the derived analytic form of the steady-state output field state, the Hong-Ou-Mandel (HOM) effect can be demonstrated by controlling the detuning frequency between the photons and the two-level systems. For more discussions on the interaction between a two-level system and two photons, interested readers may refer to \cite{Liao:10,Shi:11,Liao:13}. For the dynamics of two-level systems driven by more photons, interested readers may refer to  \cite{ZGB:10,ESS11,baragiola:2012,ZB13,Xu:15,SONG:13MULTI,CCCR:16,K16,BrodA16,Brod16,WJ17,SVF18} and the references therein.

To the best of our knowledge, the exact analytic form in the time domain of the output two-photon state for a two-level system driven by a two-photon input state has not yet been given in the literature. From a signal and system perspective, it is always desirable to have such an explicit form, as it is an important ingredient of quantum control theory and will facilitate cascade system design \cite{GJ:09,TNS:12, zhang:2014,GZ15a}. Moreover, the quantum state gives us all information about the quantum system. As demonstration, three examples are used to show that physically significant and interesting quantities can be obtained in terms of the steady-state output two-photon states derived in this paper.

Motivated by the above discussions, in this paper we derive explicit time-domain expressions of the output field states of a two-level system driven by two input photons. Two cases are studied. In Case $1$, there is one input channel which contains two photons. The analytic form of the output two-photon state in the time domain is given in Theorem \ref{thm:output}. As a by-product, the frequency-domain form is given in Corollary \ref{coro:output}. Two examples are presented, which demonstrate that different input pulses give rise to drastically different output correlations, both in the time and frequency domains. In Case $2$, there are two input channels, each of which having one photon. After deriving the analytic forms of the output two-photon state (Theorem \ref{thm:2-channel} for the time domain and Corollary \ref{coro:output1} for the frequency domain), the output two-photon time distribution and joint spectrum are simulated in Example $3$. These simulations reveal the nonlinear photon-photon interaction induced by a two-level system and the HOM effect in the two-photon scattering case.

Coherent control has been proven very effective  for controlling finite-level quantum systems.  The Hamiltonian of a finite-level quantum system usually consists of two parts:  a free Hamiltonian and a controlled Hamiltonian. The controlled Hamiltonian can be manipulated by an external field (e.g., a laser or a magnetic field) which serves as control signal. Coherent control of quantum finite-level systems concerns how to engineer the controlled Hamiltonian so that the system state can be steered in a desired manner, see. e.g.,  \cite{DD01,WW01,BCS09,LK09,BBR10,HY13,RBR18} and references therein. Essentially speaking, in all of these works, coherent control makes use of  semi-classical signals such as lasers or magnetic fields. Recently, the dynamics of a finite-level quantum system driven by one or few photons have  been studied; see the discussions in the first two paragraphs of this section. Here, we go beyond coherent control  by allowing the signals involved to be a few photons which are genuinely quantum. Due to the infinite dimensionality of the field, it is difficult to derive the explicit form of the output signal, namely the two-photon state of the field after interaction with the two-level system. In this paper, we show that the transfer function approach can be used to investigate the dynamics of a two-level system driven by two photons. Indeed, the transfer functions \eqref{transfer1} and \eqref{transfer2} and their corresponding impulse response functions \eqref{tf_2} and \eqref{tf_3}  are key to the system analysis carried out in this paper, as the steady-state output states are explicitly expressed in terms of these transfer functions;  see Theorems \ref{thm:output}, \ref{thm:2-channel} and Corollaries \ref{coro:output}, \ref{coro:output1}.  It is well-known that the transfer function approach is an important method for controller design in the classical systems and control theory, it is thus expected that the transfer functions defined in this paper will be useful for the study of controlling two-level systems driven by few photons. For instance, the transfer function approach has lately been applied to study a coherent 2-qubit feedback network, see \cite{ZP:19} for more details.

The rest of this paper is organized as follows. In Section \ref{sec:setup}, some preliminary results are reviewed, including quantum system and field, two-level system, single-photon and two-photon states. The explicit form of the output field state for a two-level system driven by a two-photon input state in a single input channel is discussed in Section \ref{sec:main}. The scenario of a two-level system driven by two counter-propagating photons is studied in Section \ref{sec:two-channel}. Section \ref{sec:con} concludes this paper.

%%%%%%%%%%%%%%%%%%%%%%%%
%%%%%%%%%%%%%%%%%%%%%%%%
%%%%%%%%%%%%%%%%%%%%%%%%
\section{Preliminary}\label{sec:setup}
\emph{Notation} $|0\rangle$ denotes the vacuum state of a free propagating field, $|g\rangle$ and $|e\rangle$ stand for the ground and excited states of a two-level system respectively. The symbol $^{\dag }$ stands for the complex conjugate of a complex number or the adjoint of a Hilbert space operator. Let
 $\sigma _{-}=|g\rangle \left\langle e\right\vert $, $\sigma_{+}=|e\rangle \left\langle g\right\vert =(\sigma _{-})^{\dag }$, and  $\sigma_{z}=2\sigma _{+}\sigma _{-}-I$, where $I$ is the identity operator. The function $\delta(t)$ is the Dirac delta. $\mathrm{i}=\sqrt{-1}$. The commutator between two operators $A$ and $B$ is $[A,B]=AB-BA$. Finally, the convolution of two functions $f(t)$ and $g(t)$ is denoted by $f\ast g (t) =\int_{-\infty}^\infty f(t-r)g(r)dr$.

%%%%%%%%%%%%%%%%%%%%%%%%
%%%%%%%%%%%%%%%%%%%%%%%%
\subsection{System and field\label{sec:sys_field}}

In this section, quantum systems and fields are briefly introduced, more details can be found in, e.g., \cite{parthasarathy:1992,gardiner:2000,james:2007,WM:08,wiseman:2010}.

The $(S,L,H)$ formalism \cite{GJ:09,TNS:12,ZJ:12} is very convenient for describing Markovian quantum systems and networks. Here, $S$ is a unitary scattering operator, the operator $L$ determines the coupling between the system and its environment (which in this paper is a light field), and the self-adjoint operator $H$ is the initial system Hamiltonian. The operators $S$, $L$, and $H$ are all  defined on the system  Hilbert space $\mathcal{H}_{S}$ in which system states reside. For clarity of presentation, in this paper we assume
that $S=I$, namely, an identity operator. The light field has a bosonic annihilation operator $b(t)$ and a creation operator $b^{\dag }(t)$; these are operators on a Fock space $\mathcal{H}_{F}$ (an infinite-dimensional Hilbert space). These field operators have the following properties%
\begin{equation}\label{ccr}\begin{aligned}
b(t)|0\rangle=0,~[b(t),b(r)]=[b^{\dag}(t),b^{\dag}(r)]=0,~[b(t),b^{\dag}(r)]=\delta(t-r),~\forall t,r\in \mathbb{R}.
\end{aligned}\end{equation}
Define integrated annihilation and creation field operators $B(t)\triangleq\int_{t_{0}}^{t}b(r)dr$ and $B^{\dag }(t)\triangleq \int_{t_{0}}^{t}b^{\dag}(r)dr$, where $t_{0}$ is the initial time, i.e., the time when the system starts its interaction with the field.

The dynamics of the joint system (system plus field) can be described by a unitary operator $U(t,t_{0})$ on the tensor product Hilbert space $\mathcal{H}_{S}\otimes \mathcal{H}_{F}$, which is the solution to the following quantum stochastic differential equation (QSDE) in It\^o form
\begin{equation}\label{U}\begin{aligned}
dU(t,t_{0})=\left\{-(L^{\dag }L/2+\mathrm{i}H)dt+LdB^{\dag}(t)-L^{\dag }dB(t)\right\} U(t,t_{0}),~~t\geq t_{0}
\end{aligned}\end{equation}
with the initial condition $U(t_{0},t_{0})=I$. In the Heisenberg picture, a system operator $X$ at time $t\geq t_{0}$ is $X(t)\equiv j_{t}(X)\triangleq U(t,t_{0})^{\dag }(X\otimes I)U(t,t_{0})$, which is an operator on $\mathcal{H}_{S}\otimes \mathcal{H}_{F}$ and solves the following QSDE
\begin{equation}\label{eq:QDSE_X}\begin{aligned}
dj_{t}(X)=j_{t}(\mathcal{L}_{00}(X))dt+j_{t}(\mathcal{L}_{01}(X))dB(t)+j_{t}(\mathcal{L}_{10}(X))dB^{\dag }(t),~~t\geq t_{0}
\end{aligned}\end{equation}
with the initial condition $j_{t_{0}}(X)=X\otimes I$, where the Evans-Parthasarathy superoperators are \cite{hudson:1984,GJN:13,SONG:13MULTI}
\[
\mathcal{L}_{00}(X) \triangleq \frac{1}{2}L^{\dag }[X,L]+\frac{1}{2}[L^{\dag },X]L-\mathrm{i}[X,H],  \ \
\mathcal{L}_{01}(X) \triangleq [L^{\dag },X],  \ \ \mathcal{L}_{10}(X) \triangleq [X,L].
\]
After interaction, the quantum output field $B_{\mathrm{out}}(t)\triangleq U(t,t_{0})^{\dag }(I\otimes B(t))U(t,t_{0})$ is generated, which is also an operator on $\mathcal{H}_{S}\otimes \mathcal{H}_{F}$ and whose dynamics are given by the following QSDE
\begin{equation}
dB_{\mathrm{out}}(t)=j_{t}(L)dt+dB(t).  \label{eq:B_out}
\end{equation}
In this paper, instead of integrated quantum processes $B(t)$ and $B_{\mathrm{out}}(t)$, we find it more convenient to work directly with the quantum processes $b(t)$ and
\begin{equation}
b_{\mathrm{out}}(t)\triangleq U(t,t_{0})^{\dag}b(t)U(t,t_{0}), \ t\geq t_{0}.
\end{equation}
Moreover, the output field annihilation operator $b_{\mathrm{out}}(t)$ enjoys the following property, (see e.g., \cite[Section 5.2]{james:2007}),
\begin{equation}
b_{\mathrm{out}}(t)=U(\tau ,t_{0})^{\dag }b(t)U(\tau ,t_{0}),\ \ \forall
\tau \geq t\geq t_{0}.  \label{eq:b_out2}
\end{equation}
Finally, let $\tau =\max \{t_{1},t_{2}\}$ for any $t_1,t_2\geq t_0$. Then by Eq. (\ref{eq:b_out2}), we have
\begin{equation}\nonumber\begin{aligned}
\left[b_{\mathrm{out}}(t_{1}),b_{\mathrm{out}}(t_{2})\right]
=U(\tau ,t_{0})^{\dag }\left[ b(t_{1}),b(t_{2})\right] U(\tau,t_{0}).
\end{aligned}\end{equation}
However, noticing Eq. (\ref{ccr}), we conclude that
\begin{equation}
\left[ b_{\mathrm{out}}(t_{1}),b_{\mathrm{out}}(t_{2})\right] =0,\ \forall
t_{1},t_{2}\geq t_{0}.  \label{eq:b_out3}
\end{equation}
Eq. (\ref{eq:b_out3}) is the so-called \textit{self-nondemolition} feature of quantum light fields \cite{james:2007}.

%%%%%%%%%%%%%%%%%%%%%%%%
%%%%%%%%%%%%%%%%%%%%%%%%
\subsection{Two-level system with a single input}\label{subsec:TLS}

In the $(S,L,H)$ formalism introduced above, the two-level system studied in this paper has the system parameters $S=I,\ L=\sqrt{\kappa }\sigma _{-},\ H=\frac{\omega_d}{2}\sigma_{z}$, where $\kappa>0$ determines the coupling strength between the system and the field, and $\omega_d\in \mathbb{R}$ is the frequency detuning (the difference between the carrier frequency of the input field and the  atomic transition frequency of the two-level system). With these parameters, by Eqs. (\ref{eq:QDSE_X})-(\ref{eq:B_out}), we have the following QSDEs
\begin{eqnarray}
\dot{\sigma}_{-}(t) &=&-\left(\frac{\kappa }{2}+\mathrm{i}\omega_d\right)
\sigma _{-}(t)+\sqrt{\kappa }\sigma _{z}(t)b(t),  \label{system_a} \\
b_{\mathrm{out}}(t) &=&\sqrt{\kappa }\sigma _{-}(t)+b(t),\ \ \ ~~t\geq t_{0}.
\label{system_b}
\end{eqnarray}
Next, we study several properties of the system (\ref{system_a})-(\ref{system_b}). Notice that
\begin{equation}\label{passive}
L|g\rangle =0,\ \ H|g\rangle =-\frac{\omega_d}{2}|g\rangle .
\end{equation}
That is, the coupling operator $L$ does not generate photons and the initial system Hamiltonian $H$ does not excite the two-level system. As a result, when this system is initialized in the vacuum state $|g\rangle$ and is driven by a two-photon state (to be discussed in Section \ref{subsec:state}), at any time instant $t$ the joint system may have either two photons in the field, or one photon in the field and one excited atomic state. That is, the number of excitations is a conserved quantity at all times. (Here the word \textquotedblleft excitation\textquotedblright\ stands for a photon or an excited two-level system.)

%Secondly, by Eq. (\ref{passive}), it can be shown that \cite[Lemma 3]{PZJ:15}, up to a global phase factor, the solution $U(t,t_{0})$ to the QSDE (\ref{U}) satisfies
%\begin{equation}
%U(t,t_{0})\left\vert 0g\right\rangle =\left\vert 0g\right\rangle .
%\label{eq:passive}
%\end{equation}
%Multiplying both sides of Eq. (\ref{eq:passive}) by $U(t,t_{0})^{\dag }$ yields
%\begin{equation}
%\left\vert 0g\right\rangle =U(t,t_{0})^{\dag }\left\vert 0g\right\rangle .
%\label{eq:passive3}
%\end{equation}
%Conjugating both sides of Eq. (\ref{eq:passive3}) we get%
%\begin{equation}
%\left\langle 0g\right\vert U(t,t_{0})=\left\langle 0g\right\vert .
%\label{eq:passive2}
%\end{equation}
%Furthermore, by Eqs. (\ref{eq:passive}) and (\ref{eq:passive3}) we have
%%\begin{equation}\label{feb27_1}\begin{aligned}
%%\sigma_{z}(t)|0g\rangle=U(t,t_{0})^{\dag }\sigma_{z}(t_{0})U(t,t_{0})|0g\rangle
%%=U(t,t_{0})^{\dag }\sigma_{z}(t_{0})|0g\rangle
%%=-U(t,t_{0})^{\dag }|0g\rangle
%%=-|0g\rangle.
%%\end{aligned}\end{equation}
%%That is,
%\begin{equation}
%\sigma _{z}(t)|0g\rangle =-|0g\rangle ,\ \ \ \left\langle 0g\right\vert
%\sigma _{z}(t)=-\left\langle 0g\right\vert .  \label{eq:feb23_temp2}
%\end{equation}
%Thirdly,
It is worth mentioning the quantum causality conditions \cite{Xu:15}
\begin{eqnarray}
\lbrack X(t),b(\tau )]=&[X(t),b^{\dagger }(\tau )]=0,\ \ t\leq \tau,\label{causality_a}\\
\lbrack X(t),b_{\mathrm{out}}(\tau )]=&[X(t),b_{\mathrm{out}}^{\dagger}(\tau )]=0,\ \ t\geq \tau.\label{causality_b}
\end{eqnarray}
Eq. (\ref{causality_a}) indicates that the system operator $X(t)$ is influenced by the \textit{past} input field $b(r)$ ($t_{0}\leq r<t$). On the other hand, Eq. (\ref{causality_b}) tells us that the past output field is not affected by the current and future system state.
%On the other hand, it is clear that $[X(t),b_{\mathrm{out}}(\tau )]=0$ for $t>\tau$.
%We show that Eq. (\ref{causality_b}) is valid when $t=\tau$,
%\begin{equation}\nonumber\begin{aligned}
%\lbrack X(t),b_{\mathrm{out}}(t)]=U(t,t_{0})^{\dag}\left[ X\otimes I,I\otimes b(t)\right]U(t,t_{0})
%=0,\\
%\lbrack X(t),b_{\mathrm{out}}^{\dagger }(t )]=U(t,t_{0})^{\dag}\left[ X\otimes I,I\otimes b^{\dag }(t)\right]U(t,t_{0})
%=0.
%\end{aligned}\end{equation}
Finally, because of Eq. (\ref{causality_a}), $\sigma _{z}(t)b(t)$\ in Eq. (\ref{system_a}) is equal to $b(t)\sigma _{z}(t)$. Moreover, $\sigma_z(t)\ket{0g}=-\ket{0g}$. Therefore, post-multiplying both sides of Eqs. \eqref{system_a}-\eqref{system_b}  by $\ket{0g}$ yields a linear dynamical system. In this sense, we can define the impulse response function
\begin{equation}\label{tf_2}
g_{G}(t)\triangleq \left\{
\begin{array}{cc}
\delta (t)-\kappa e^{-(\frac{\kappa }{2}+\mathrm{i}\omega_d )t}, & t\geq 0, \\
0, & t<0,
\end{array}
\right.
\end{equation}
and the corresponding  transfer function
\begin{equation}\label{transfer1}
G[s]=\frac{s+\mathrm{i}\omega_d-\frac{\kappa}{2}}{s+\mathrm{i}\omega_d+\frac{\kappa}{2}}.
\end{equation}

\begin{remark}
It turns out that $g_{G}(t)$ and $G[s]$ are very helpful  in presenting the analytic forms of the two-photon output field state of a two-level system driven by a two-photon input state; see Theorem \ref{thm:output} and Corollary \ref{coro:output} for details.
\end{remark}

%%%%%%%%%%%%%%%%%%%%%%%%
%%%%%%%%%%%%%%%%%%%%%%%%
\subsection{Two-photon states}\label{subsec:state}

Compared with Gaussian states, single- and multi- photon states are highly non-classical and have found promising applications in  quantum computation and quantum signal processing \cite{LMS15,Nysteen15,Roulet16,gu2017microwave,lodahl2017chiral,RWF17}. Given a function $\xi \in L_{2}(\mathbb{R},\mathbb{C})$, define an operator
\begin{equation}\label{B_xi}
B(\xi )\triangleq \int_{-\infty }^{\infty }\xi ^{\dag }(t)b(t)dt,
\end{equation}
whose adjoint operator is
\begin{equation}
B^{\dag }(\xi )=\int_{-\infty }^{\infty }\xi (t)b^{\dag }(t)dt.
\label{eq:B^ast_xi}
\end{equation}
A continuous-mode single-photon state can be defined as
\begin{equation}
|1_{\xi }\rangle \triangleq B^{\dag }(\xi )|0\rangle ,  \label{1_photon}
\end{equation}
where $\Vert \xi \Vert =1$ for normalization. For example, if the pulse shape is $\xi(t)=-\sqrt{\gamma}e^{\frac{\gamma}{2}t}(1-u(t))$, where $u(t)$ is the Heaviside function
\begin{equation}\label{Hea}
u(t) =
\bigg\{
\begin{array}{cc}
1, & t>0,\\
0, & t\leq 0,
\end{array}
\end{equation}
then in the frequency domain  we have $f[\omega] \triangleq \int_{-\infty}^\infty e^{-\mathrm{i}\omega t} \xi(t)dt = \frac{1}{\sqrt{2\pi}} \frac{\sqrt{\gamma}}{\mathrm{i}\omega-\gamma/2}$.
Clearly, $f[\omega]$ describes a Lorentzian spectrum with FWHM $\gamma$.
In the calculation of various few-photon states, the notation
\begin{equation}
|1_{t}\rangle \triangleq b^{\dag }(t)|0\rangle ,\ \ \forall t\in \mathbb{R},
\label{1t}
\end{equation}
turns out to be very useful.  Roughly speaking, $1_{t}$ means that a photon is generated by $b^{\dag }(t)$  from the vacuum.  By Eq. (\ref{ccr}), we have $\left\langle 1_{t}|1_{r}\right\rangle=\delta(t-r)$.
Moreover, by Eqs. (\ref{eq:B^ast_xi}) and (\ref{1t}), the single-photon state $|1_{\xi }\rangle $ can be re-written as
\begin{equation*}
|1_{\xi }\rangle =\int_{-\infty }^{\infty }\xi(t)|1_{t}\rangle dt.
\end{equation*}
That is, the single-photon state $|1_{\xi }\rangle $ is in the form of continuum superposition of $|1_{t}\rangle $. Consequently, $\left\{|1_{t}\rangle :t\in \mathbb{R}\right\}$ is a complete single-photon basis.
Similarly
\begin{equation}
\int_{-\infty }^{\infty }dl\ \left\vert 1_{l}g\right\rangle \left\langle
1_{l}g\right\vert +|0e\rangle \langle 0e|  \label{id_1}
\end{equation}
is an identity operator in the one-excitation case.

In what follows, we introduce two-photon states. Given two functions $\xi_{1},\xi _{2}\in L_{2}(\mathbb{R},\mathbb{C})$ satisfying $\Vert \xi_{1}\Vert =\Vert \xi _{2}\Vert =1$, we may define the following two-photon state
\begin{equation}\label{state_2}
\ket{2_{\xi_1, \xi_2 }} \triangleq \frac{1}{\sqrt{N_{2}}}B^{\dag }(\xi _{1})B^{\dag}(\xi _{2})|0\rangle,
\end{equation}
where $N_{2}=1+|\langle \xi _{1}|\xi _{2}\rangle |^{2}$ is the normalization coefficient. If $\xi _{1}\equiv \xi _{2}$, then $|2_{\xi_1,\xi_2}\rangle$ is a continuous-mode two-photon Fock state \cite{baragiola:2012, SONG:13MULTI}. More generally,  an arbitrary continuous-mode two-photon state is given by
\[
\int_{-\infty }^{\infty }dp_{1}\int_{-\infty }^{\infty }dp_{2}\
f(p_{1},p_{2})b^{\dag }(p_{1})b^{\dag }(p_{2})\left\vert 0\right\rangle
=\int_{-\infty }^{\infty }dp_{1}\int_{-\infty }^{\infty }dp_{2}\
f(p_{1},p_{2})\left\vert 1_{p_{1}}1_{p_{2}}\right\rangle,
\]
where $f(p_{1},p_{2})$ is an ordinary function of time variables $p_1$ and $p_2$, satisfying the symmetry property $f(p_1,p_2)=f(p_2,p_1)$. It can be easily checked that
\begin{equation}\nonumber\begin{aligned}
\int_{-\infty}^{\infty}dp^\prime_{1}\int_{-\infty}^{\infty}dp^\prime_{2}\int_{-\infty}^{\infty}dp_{1}\int_{-\infty}^{\infty}dp_{2}
f^\dag(p^\prime_{1},p^\prime_{2})f(p_{1},p_{2})\langle1_{p^\prime_1}1_{p^\prime_2}\vert1_{p_1}1_{p_2}\rangle
=2.
\end{aligned}\end{equation}
Therefore, $\left\{ \frac{1}{\sqrt{2}}\left\vert 1_{p_{1}}1_{p_{2}}\right\rangle :p_{1},p_{2}\in \mathbb{R}\right\}$ is a complete orthonormal basis of continuous-mode two-photon pure states. As a result,
\begin{equation}\label{id_2}\begin{aligned}
\frac{1}{2}\int_{-\infty}^{\infty}dp_1\int_{-\infty}^{\infty}dp_2\left\vert1_{p_1}1_{p_2}g\right\rangle\left\langle1_{p_1}1_{p_2}g\right\vert
+\int_{-\infty}^{\infty}dp\left\vert1_{p}e\right\rangle\left\langle1_pe\right\vert
\end{aligned}\end{equation}
is an identity operator for the 2-excitation composite system.

%%%%%%%%%%%%%%%%%%%%%%%
%%%%%%%%%%%%%%%%%%%%%%%
%%%%%%%%%%%%%%%%%%%%%%%
\section{One-channel case}\label{sec:main}

In this section, we study the dynamics of a two-level system which is driven by a two-photon state $|2_{\xi_1,\xi_2}\rangle$. The main results are explicit expressions of the steady-state output field state in the time and frequency domains.

\subsection{The output field state in the time domain}\label{sec:main-time} In this subsection, an analytic form of the output two-photon state is presented in the time domain.

Integrating (\ref{system_a}) from $t_{0}$ to $t$ gives
\begin{eqnarray}
\sigma_-(t)&=&e^{-\left( \frac{\kappa }{2}+\mathrm{i}\omega_d \right)(t-t_{0})}\sigma _{-}(t_{0})
+\sqrt{\kappa }\int_{t_{0}}^{t}dr\ e^{-\left(\frac{\kappa }{2}+\mathrm{i}\omega_d \right) (t-r)}\sigma _{z}(r)b(r), \\\label{sigma_minus}
\ b_{\mathrm{out}}(t)&=&\sqrt{\kappa }\sigma _{-}(t)+b(t).  \label{output}
\end{eqnarray}
Assume that the two-level system is initialized in the ground state $|g\rangle$ and the input field is in the two-photon state $|2_{\xi_1,\xi_2}\rangle$. Then the initial joint system-field state is
\begin{equation}
\left\vert \Psi (t_{0})\right\rangle =|2_{\xi_1,\xi_2}g\rangle =\frac{1}{\sqrt{N_{2}
}}B^{\dag }(\xi _{1})B^{\dag }(\xi _{2})\left\vert 0g\right\rangle .
\label{eq:initial_state}
\end{equation}
By the Schr\"{o}dinger equation, the joint system-field state at time $t\geq t_{0}$ is
\begin{equation}
\left\vert \Psi (t)\right\rangle =U(t,t_{0})\left\vert \Psi
(t_{0})\right\rangle .  \label{state}
\end{equation}

In this paper, we are interested in the steady-state output field state, i.e., we assume that the interaction starts in the remote past ($t_{0}=-\infty$) and terminates in the far future ($t=\infty $), \cite{SF:05,Shen:07,FKS:10,Shi:11,Rep:12,Re:12,Xu:13,LP:14,Xu:15,PZJ:15,PZJ:16}. In the steady state, the two-level system is in the ground state $\left\vert g\right\rangle $ and the two photons are in the output field. Thus, the steady-state output field state is obtained by tracing out the system
\begin{equation}
\left\vert \Psi _{\mathrm{out}}\right\rangle =\lim_{\begin{subarray}{c}t_0\rightarrow-\infty \\ t\rightarrow\infty\end{subarray}}\left\langle g|\Psi (t)\right\rangle =\lim_{\begin{subarray}{c}t_0\rightarrow-\infty \\ t\rightarrow\infty\end{subarray}}\left\langle g|U(t,t_{0})| \Psi
(t_{0})\right\rangle .
\label{eq:feb20_1}
\end{equation}
The aim of this section is to derive analytic expressions of $\left\vert \Psi_{\mathrm{out}}\right\rangle$. Substituting Eq. (\ref{eq:initial_state}) into Eq. (\ref{eq:feb20_1}) yields
\begin{equation}\label{eq:output}\begin{aligned}
\left\vert\Psi_{\mathrm{out}}\right\rangle
=\frac{1}{\sqrt{N_{2}}}\lim_{\begin{subarray}{c}t_0\rightarrow-\infty \\ t\rightarrow\infty\end{subarray}}\int_{t_{0}}^{t}dt_{1}\xi_{1}(t_{1})\int_{t_{0}}^{t}dt_{2}\xi_{2}(t_{2})
\left\langle g\right\vert U(t,t_{0})b^{\dag }(t_{1})b^{\dag
}(t_{2})\left\vert 0g\right\rangle.
\end{aligned}\end{equation}

%Also, another function $\eta(t)$ via the convolution is defined as
%\begin{equation}\label{eta}
%\eta(t)\triangleq g_{G}\ast\xi(t)\equiv\int_{-\infty}^{\infty}g_{G}(t-r)\xi(r)dr.
%\end{equation}
%It has been proven in \cite{PZJ:15} that the steady-state output field state of a two-level system driven by a single-photon state is another single-photon state. More specifically, we state the following lemma borrowed from \cite{PZJ:15}.
%\begin{lemma}
%(\cite[Theorem 4]{PZJ:15}) \label{lem:1-photon} If the two-level system (\ref{system_a})-(\ref{system_b}) is initialized in the ground state $|g\rangle$ and is driven by a single-photon state $|1_{\xi }\rangle $ defined in Eq. (\ref{1_photon}), then the steady-state output field state  is another single-photon state $|1_{\eta }\rangle $ with the pulse shape $\eta(t)$ given in Eq. (\ref{eta}).
%\end{lemma}

%The following lemma is important in the proof of the subsequent lemmas.
%
%\begin{lemma}\label{lem:basis}
%If the system (\ref{system_a})-(\ref{system_b}) is initialized in the ground state $\left\vert g\right\rangle $ and driven by a two-photon state $|2_{\xi_1,\xi_2}\rangle$, then
%\begin{equation}\label{basis_2}\begin{aligned}
%&\left\langle g\right\vert U(t,t_{0})b^{\dag }(t_{1})b^{\dag}(t_{2})\left\vert 0g\right\rangle\\
%=&\frac{1}{2}\int_{-\infty }^{\infty}dp_{1}\int_{-\infty }^{\infty }dp_{2}\ \left\vert1_{p_{1}}1_{p_{2}}\right\rangle
%\left\langle 1_{p_{1}}1_{p_{2}}g\right\vert
%U(t,t_{0})b^{\dag }(t_{1})b^{\dag }(t_{2})\left\vert 0g\right\rangle , \ t\geq t_{0}.
%\end{aligned}\end{equation}
%\end{lemma}

%\textbf{Proof}.
As discussed above, the system (\ref{system_a})-(\ref{system_b}) satisfies the conditions (\ref{passive}). Hence, if the system is initialized in the ground state $\left\vert g\right\rangle $ and driven by a two-photon state $|2_{\xi_1,\xi_2}\rangle $, the number of excitations of the joint system is always two for all times. Consequently, by using the identity operator in Eq. \eqref{id_2}, we have
\begin{equation}\label{basis_2}\begin{aligned}
&\left\langle g\right\vert U(t,t_{0})b^{\dag }(t_{1})b^{\dag}(t_{2})\left\vert 0g\right\rangle\\
=&\frac{1}{2}\int_{-\infty }^{\infty }dp_{1}\int_{-\infty }^{\infty}dp_{2}\ \left\vert 1_{p_{1}}1_{p_{2}}\right\rangle
\left\langle1_{p_{1}}1_{p_{2}}g\right\vert U(t,t_{0})b^{\dag }(t_{1})b^{\dag}(t_{2})\left\vert 0g\right\rangle, \ t\geq t_0.
\end{aligned}\end{equation}
%$\square$

\begin{remark}
The term $\left\langle g\right\vert U(t,t_{0})b^{\dag }(t_{1})b^{\dag}(t_{2})\left\vert 0g\right\rangle $ in Eq. (\ref{basis_2}) is an (unnormalized) two-photon field state after tracing out the system. Eq. (\ref{basis_2}) expresses this field state in terms of a coherent superposition of a complete two-photon basis $$\left\{\frac{1}{\sqrt{2}}\left\vert
1_{p_{1}}1_{p_{2}}\right\rangle :p_{1},p_{2}\in\mathbb{R}\right\}$$ with weights $\frac{1}{\sqrt{2}}\left\langle 1_{p_{1}}1_{p_{2}}g\right\vert
U(t,t_{0})b^{\dag }(t_{1})b^{\dag }(t_{2})\left\vert 0g\right\rangle$.
\end{remark}

The substitution of Eq. (\ref{basis_2}) into Eq. (\ref{eq:output}) produces
\begin{align}
\left\vert \Psi _{\mathrm{out}}\right\rangle
=&\; \frac{1}{2\sqrt{N_{2}}}\int_{-\infty }^{\infty }dp_{1}\int_{-\infty}^{\infty }dp_{2}\left\vert 1_{p_{1}}1_{p_{2}}\right\rangle \nonumber
\\
&\times\lim_{\begin{subarray}{c}t_0\rightarrow-\infty \\ t\rightarrow\infty\end{subarray}}\int_{t_0}^{t}dt_1\; \xi _{1}(t_1)\int_{t_{0}}^{t}dt_2\; \xi_2(t_2)\left\langle1_{p_1}1_{p_2}g\right\vert U(t,t_{0})b^{\dag}(t_1)b^{\dag }(t_2)\left\vert 0g\right\rangle.
\label{eq:feb22_1}
\end{align}
Therefore, in order to get an analytic form of $\left\vert\Psi_{\mathrm{out}}\right\rangle$, we will have to calculate the term in Eq. \eqref{eq:feb22_1}.  The calculations are given in Appendix \ref{appendixa}; see Lemma \ref{lem:out4}.

The following result presents an analytic form of  $\left\vert\Psi_{\mathrm{out}}\right\rangle$ in the time domain.

%\begin{lemma}\label{lem:out_1}
%The steady-state output field state $\left\vert \Psi_{\mathrm{out}}\right\rangle $ in Eq. (\ref{eq:feb20_1}) can be re-written as
%\begin{equation}\label{eq:feb23_output}
%\left\vert\Psi _{\mathrm{out}}\right\rangle \\
%= \frac{1}{2\sqrt{N_{2}}}\int_{-\infty }^{\infty }dp_{1}\int_{-\infty}^{\infty }dp_{2}\ \Gamma(p_1,p_2)  \left\vert 1_{p_{1}}1_{p_{2}}\right\rangle,
%\end{equation}
%where
%\begin{align*}
%\Gamma(p_1,p_2)&=\lim_{t_{0}\rightarrow -\infty ,t\rightarrow \infty }\int_{t_{0}}^{t}dt_{1}\
%\xi _{1}(t_{1})\int_{t_{0}}^{t}dt_{2}\ \xi _{2}(t_{2})\times \\
%&\Big[\sqrt{\kappa }\left\langle 0g|\sigma _{-}(p_{1})b_{\mathrm{out}}(p_{2})b^{\dag }(t_{1})b^{\dag }(t_{2})|0g\right\rangle
%+\left\langle0g|b(p_{1})b_{\mathrm{out}}(p_{2})b^{\dag }(t_{1})b^{\dag}(t_{2})|0g\right\rangle\Big].
%\end{align*}
%\end{lemma}
%The proof of Lemma \ref{lem:out_1} is given in Appendix \ref{appendixa0}.
%
%Lemma \ref{lem:out_1} tells us that to derive the steady-state output field state $\left\vert \Psi _{\mathrm{out}}\right\rangle $, we have to calculate
%the two terms on the right-hand side of Eq. (\ref{eq:feb23_output}), namely, $\left\langle 0g|\sigma _{-}(p_{1})b_{\mathrm{out}}(p_{2})b^{\dag }(t_{1})b^{\dag }(t_{2})|0g\right\rangle$, and $ \left\langle0g|b(p_{1})b_{\mathrm{out}}(p_{2})b^{\dag }(t_{1})b^{\dag}(t_{2})|0g\right\rangle$. {\color{red} Such expressions are omitted and provided in Appendix \ref{appendixb}, as the proofs of them are straightforward.}

\begin{theorem}\label{thm:output}
The steady-state output field state $\left\vert\Psi_{\mathrm{out}}\right\rangle$ is
\begin{equation}\label{output_ss_2}\begin{aligned}
\left\vert\Psi_{\mathrm{out}}\right\rangle
=\frac{1}{2\sqrt{N_2}}\int_{-\infty}^{\infty}dp_1\int_{-\infty}^{\infty}dp_2 \; \eta(p_1,p_2)
b^{\dag}(p_1)b^{\dag}(p_2)\left\vert0\right\rangle,
\end{aligned}\end{equation}
where
\begin{equation}\label{eta_may27}
\eta(p_1,p_2)=\nu_1(p_1)\nu_2(p_2)+\nu_1(p_2)\nu_2(p_1)+\zeta(p_1,p_2)+\zeta(p_2,p_1),
\end{equation}
with
\begin{equation}\label{eta_9}
\nu_j (t) = g_G\ast\xi_j (t),\ \ j=1,2,
\end{equation}
and
\begin{equation}\label{zeta_1}\begin{aligned}
\zeta(p_1,p_2)
= &\; 2\kappa\ e^{-\frac{\kappa}{2}(p_1-p_2)-\mathrm{i}\omega_d(p_1+p_2)}\int_{p_2}^{p_1}d\tau\ e^{2\mathrm{i}\omega_d\tau}\\
&\times\left[\xi_1(\tau)\xi_2(\tau)-\frac{\xi_1(\tau)\nu_2(\tau)+\nu_1(\tau)\xi_2(\tau)}{2}\right],~~p_1\geq p_2.
\end{aligned}\end{equation}
In particular, if $\xi _{1}=\xi _{2}=\xi$ (an input two-photon Fock state) and  $\omega_d =0$, then $\nu_1=\nu_2=\nu$, and
\begin{equation}\nonumber\begin{aligned}
\zeta(p_1,p_2)=
\bigg\{
\begin{array}{cc}
2\kappa\ e^{-\frac{\kappa}{2}(p_1-p_2)}\int_{p_2}^{p_1}d\tau\ \xi(\tau)\left[\xi(\tau)-\nu(\tau)\right], &  p_1\geq p_2,\\
0, & p_1<p_2.
\end{array}
\end{aligned}\end{equation}
\end{theorem}

The proof of Theorem \ref{thm:output} is given in Appendix \ref{appendixa}.

%\begin{remark}
%By the explicit form of $\zeta(p_1,p_2)$ which is given in Eq. (\ref{zeta_1}), we can see that the interaction between the two-level system and the two input photons is not a linear transformation.
%\end{remark}

%%%%%%%%%%%%%%%%%%%%%%%%
%%%%%%%%%%%%%%%%%%%%%%%%
\subsection{The output field state in the frequency domain}\label{sec:main-fre}

In this subsection, an analytic form of the output two-photon state is presented in the frequency domain.

The Fourier transform of the annihilation operator $b(t)$ is defined as
\begin{equation}
b[\omega]=\frac{1}{\sqrt{2\pi}}\int_{-\infty}^{\infty}e^{-\mathrm{i}\omega t}b(t)dt,~~\omega\in\mathbb{R}.
\end{equation}
Similarly, the Fourier transform of the function $\xi(t)$ in Eq. (\ref{B_xi}) can be defined as
\begin{equation}
\xi[\mu]=\frac{1}{\sqrt{2\pi}}\int_{-\infty}^{\infty}e^{-\mathrm{i}\mu t}\xi(t)dt,~~\mu\in\mathbb{R}.
\end{equation}
It can be easily verified that
\[
\int_{-\infty}^{\infty}\xi(t)b^\dagger(t)dt=\int_{-\infty}^{\infty}\xi[\mu]b^\dagger[\mu]d\mu.
\]
%then the steady-state output field state $|\Psi_{\mathrm{out}}\rangle$ in Eq. (\ref{eq:feb23_output}) becomes
%\begin{equation}\label{May23}\begin{aligned}
%|\Psi_{\mathrm{out}}\rangle
%=&\frac{1}{2\sqrt{N_2}}\int_{-\infty}^{\infty}dp_1\int_{-\infty}^{\infty}dp_2|1_{p_1}1_{p_2}\rangle
%\int_{-\infty}^{\infty}d\mu_1\int_{-\infty}^{\infty}d\mu_2\xi_1[\mu_1]\xi_2[\mu_2]\\
%&\times\langle0g|b_{\mathrm{out}}(p_1)b_{\mathrm{out}}(p_2)b^\dagger[\mu_1]b^\dagger[\mu_2]|0g\rangle.
%\end{aligned}\end{equation}
%Finally,
Fourier transforming $|\Psi_{\mathrm{out}}\rangle$ in Eq. \eqref{output_ss_2} with respect to the time variables $p_1$ and $p_2$, yields an analytic expression of $|\Psi_{\mathrm{out}}\rangle$ in the frequency domain, which is given by the following result.

\begin{corollary}\label{coro:output}
The steady-state output field state $\left\vert\Psi_{\mathrm{out}}\right\rangle$ in the frequency domain is
\begin{equation}\label{May23b}\begin{aligned}
\left\vert\Psi_{\mathrm{out}}\right\rangle
=&\frac{1}{2\sqrt{N_2}}\int_{-\infty}^{\infty}d\omega_1\int_{-\infty}^{\infty}d\omega_2\
\eta[\omega_1,\omega_2]b^\dagger[\omega_1]b^\dagger[\omega_2]|0\rangle,
\end{aligned}\end{equation}
where
\begin{equation}\label{May23c}\begin{aligned}
\eta[\omega_1,\omega_2]
=&\; G[\mathrm{i}\omega_1]G[\mathrm{i}\omega_2]\left(\xi_1[\omega_2]\xi_2[\omega_1]+\xi_1[\omega_1]\xi_2[\omega_2]\right) \\
&+\frac{1}{\pi\kappa}\int_{-\infty}^{\infty}d\mu_1\ \xi_1[\mu_1]\xi_2[\omega_1+\omega_2-\mu_1]g(\omega_1,\omega_2,\mu_1,\omega_1+\omega_2-\mu_1),
\end{aligned}\end{equation}
and
\begin{equation}\label{May23d}
g(\omega_1,\omega_2,\mu_1,\mu_2)=(G[\mathrm{i}\omega_1]-1)(G[\mathrm{i}\omega_2]-1)(G[\mathrm{i}\mu_1]+G[\mathrm{i}\mu_2]-2),
\end{equation}
with $G[s]$ given in Eq. (\ref{transfer1}). In particular, if $\xi_1=\xi_2=\xi$ (an input two-photon Fock state) and $\omega_d=0$, then
\begin{equation}\label{May23e}\begin{aligned}
\eta[\omega_1,\omega_2]
=&\; 2G[\mathrm{i}\omega_1]G[\mathrm{i}\omega_2]\xi[\omega_1]\xi[\omega_2] \\
&+\frac{1}{\pi\kappa}\int_{-\infty}^{\infty}d\mu_1\ \xi[\mu_1]\xi[\omega_1+\omega_2-\mu_1]g(\omega_1,\omega_2,\mu_1,\omega_1+\omega_2-\mu_1).
\end{aligned}\end{equation}
\end{corollary}

%\begin{remark}
%The term $\langle0g|b_{\mathrm{out}}[\omega_1]b_{\mathrm{out}}[\omega_2]b^\dagger[\mu_1]b^\dagger[\mu_2]|0g\rangle$ in Eq. (\ref{May23a}) is exactly the $S$-matrix in the input-output formalism, which has been well studied in, e.g., \cite{Shen:07,FKS:10,Shi:11,Re:12,Xu:13,Xu:15}.
%\end{remark}

\subsection{Numerical examples}\label{subsec:33} In this section,  two examples are used to illustrate Theorem \ref{thm:output} (for the time domain) and Corollary \ref{coro:output} (for the frequency domain). In Example 1, the input photons are assumed to have Gaussian pulse shapes. In Example 2,   the input photons are assumed to have rising exponential pulse shapes. Simulations show that different input pulse shapes have a remarkable influence on the  probability distributions and joint spectra of output photons.

\begin{figure}
\begin{center}
\includegraphics[scale=0.4]{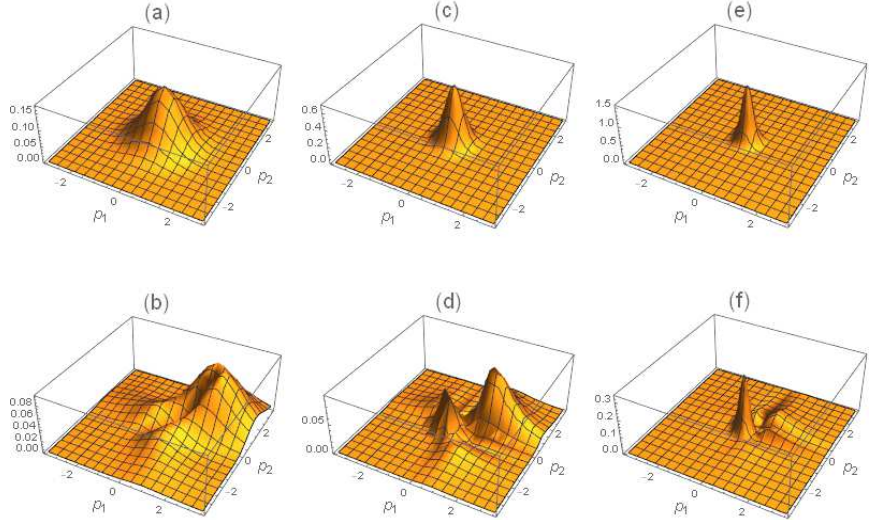}
\caption{(Color online) The input two-photon  probability distributions $\frac{1}{2}|\xi(p_1)\xi(p_2)|^2$   and output two-photon probability distributions $\frac{1}{8}|\eta(p_1,p_2)|^2$ for different bandwidths $\Omega$ of the input Gaussian pulses. (a), (c), and (e) are the input two-photon probability distributions with $\Omega=1.46\kappa$, $\Omega=2.92\kappa$, and $\Omega=4.38\kappa$, respectively. (b), (d), and (f) are the corresponding output two-photon probability distributions.}
\label{pulse}
\end{center}
\end{figure}

\emph{Example} 1. In this example, we firstly study the input and output two-photon probability distributions in the time domain. Simulations results are given in Fig. \ref{pulse}.

In Fig. \ref{pulse}, we consider that the two-photon input Fock state is with Gaussian pulse shape, i.e., the two-photon wave packets are given by
\begin{equation}\label{gaussian}
\xi_1(t)=\xi_2(t)=\left(\frac{\Omega^2}{2\pi}\right)^{\frac{1}{4}}\exp\left(-\frac{\Omega^2}{4}t^2\right),
\end{equation}
where $\Omega$ is the photon frequency bandwidth. For the scenario of a two-level system driven by a single-photon Gaussian state, it is clear to see that the output single-photon state is no longer of Gaussian pulse shape. Moreover, the excitation probability attains the maximum value $0.8$ when the photon bandwidth is chosen to be $\Omega=1.46\kappa$; see \cite{GJNC:12,SONG:13MULTI} for more details.

Here, for the two-photon scenario considered in Fig. \ref{pulse}, the input two-photon bandwidths are chosen to be (a) $\Omega=1.46\kappa$, (c) $\Omega=2.92\kappa$, and (e) $\Omega=4.38\kappa$, respectively. The corresponding output two-photon probability distributions are given by (b), (d), and (f). By comparing these subfigures, it can be observed that the output two photons are with non-Gaussian pulse shapes and their probability distributions in the time domain are more spread out than their input counterparts. Moreover, if the photon bandwidth is set to be $\Omega=2.92\kappa$, the output two-photon probability distribution consists of two peaks, each of which is similar to the input probability distribution. Interestingly, it has been shown in \cite{SONG:13MULTI} that $\Omega=2.92\kappa$ is exactly the optimal ratio for the atomic excitation by two input Gaussian photons.

\begin{figure}
\begin{center}
\includegraphics[scale=0.3]{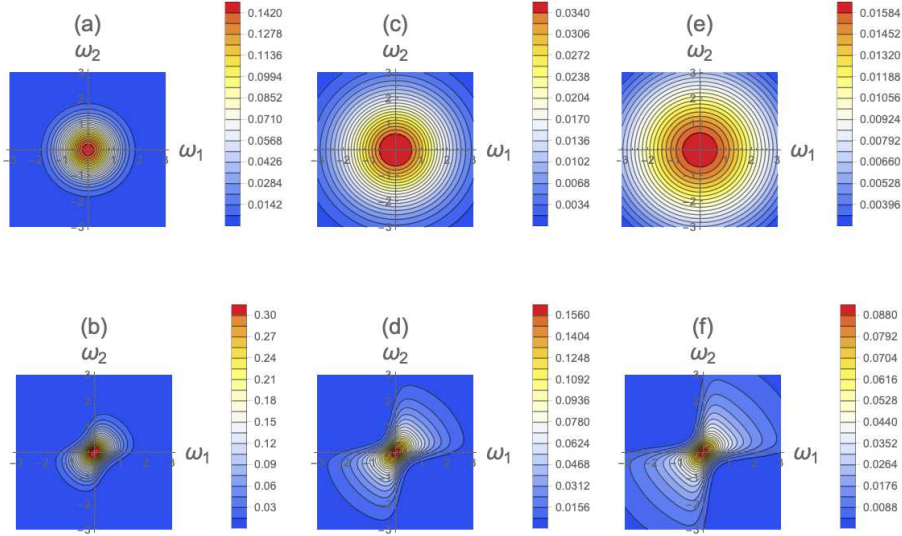}
\caption{(Color online) The input two-photon joint spectra $\frac{1}{2}|\xi[\omega_1]\xi[\omega_2]|^2$ and output two-photon joint spectra $\frac{1}{8}|\eta[\omega_1,\omega_2)|^2$ for different bandwidths $\Omega$. (a), (c), and (e) are the input two-photon joint spectra with $\Omega=1.46\kappa$, $\Omega=2.92\kappa$, and $\Omega=4.38\kappa$, respectively. (b), (d), and (f) are the corresponding output two-photon joint spectra.}
\label{Gaussian_fre}
\end{center}
\end{figure}

Next, we study the input and output two-photon joint spectra, see Fig. \ref{Gaussian_fre}. By comparing Fig. \ref{Gaussian_fre}(a), (c), (e) and Fig. \ref{Gaussian_fre}(b), (d), (f) we see that in the frequency domain the output photons are more concentrated at the origin than their input counterparts. Moreover, comparing Figs. \ref{pulse} and \ref{Gaussian_fre} we see that the scaling $\Omega=2.92\kappa$ gives rise to more interesting phenomenon in the time domain than in the frequency domain. Therefore, the output two-photon state can be understood much better when it is viewed from {\it both} the time and frequency domains.

%%%%%%%%%%%%%%%%%%%%%%%%%%
\emph{Example} 2. Let the input two-photon state be $\ket{2_{\xi,\xi}}$, where $\xi(t)=-\sqrt{\gamma}e^{\frac{\gamma}{2}t}(1-u(t))$ with $u(t)$ being the Heaviside function in Eq. \eqref{Hea}. Also, we fix $\gamma=0.1$.
% the pulse shape wave packet is of Lorentzian spectrum, i.e.,
%\begin{equation}\begin{aligned}\label{Lor}
%\xi_1[\omega]=\xi_2[\omega]=\frac{\sqrt{\gamma}}{\sqrt{2\pi}(\mathrm{i}\omega-\frac{\gamma}{2})},
%\end{aligned}\end{equation}
%where $\gamma$ is the full width at half maximum (FWHM). Here, we fix $\gamma=0.1$.

\begin{figure}
\begin{center}
\includegraphics[scale=0.4]{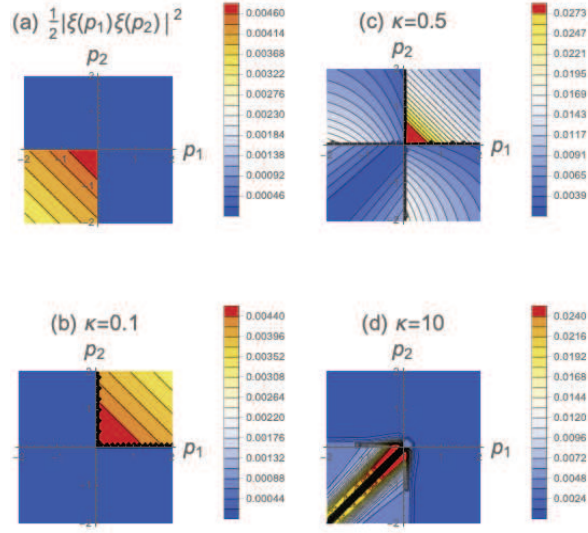}
\caption{(Color online) The input (a) and output two-photon probability distributions for different couplings: (b) $\kappa= 0.1$; (c) $\kappa = 0.5$; and (d) $\kappa = 10$.}
\label{Lortime}
\end{center}
\end{figure}

By means of Theorem \ref{thm:output}, the input and output two-photon probability distributions are plotted in Fig. \ref{Lortime}. Interestingly, when the coupling $\kappa=\gamma=0.1$, the output two-photon probability distribution is almost symmetric with that of the input, cf. Fig. \ref{Lortime}(a) and (b). When $\kappa=0.5$ in Fig. \ref{Lortime}(c), the output two photons can be distributed in all regions. However, when the coupling is relatively large ($\kappa=10$ in contrast to $\gamma=0.1$), the output two photons are mainly distributed in the region $p_1,p_2\leq0$ as shown in Fig. \ref{Lortime}(d).

\begin{figure}
\begin{center}
\includegraphics[scale=0.4]{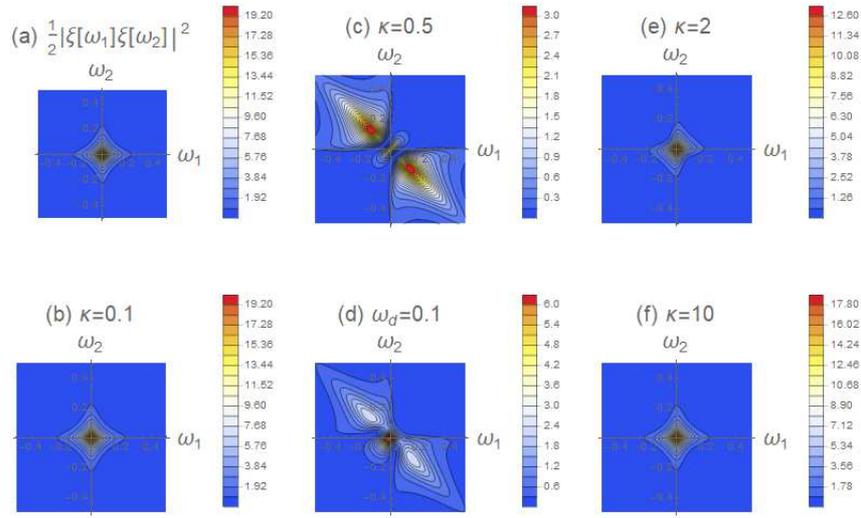}
\caption{(Color online) The input and output two-photon joint spectra. The input two-photon joint spectrum is plotted in (a). The output two-photon joint spectra with different couplings are given in: (b) $\kappa=0.1$; (c) $\kappa=0.5$; (e) $\kappa=2$; and (f) $\kappa=10$. (d) corresponds  to the nonzero detuning  case ($\omega_d=0.1, \kappa=0.5$). There is a red point at the origin in (b), (e) and (f), where the maximal value of $\frac{1}{8}|\eta[\omega_1,\omega_2]|^2$ is attained. In contrast, there are two maximal values in (c), which are along the line $\omega_1+\omega_2=0$, thus indicating photon-photon anti-correlation.}
\label{Lorfre}
\end{center}
\end{figure}

In what follows, we discuss the correlation between the two output photons in the frequency domain. Based on Corollary \ref{coro:output},  the output two-photon joint spectra $\frac{1}{8}|\eta[\omega_1,\omega_2]|^2$  are plotted in  Fig. \ref{Lorfre}.  It can be observed that the output two-photon joint spectra are almost  the same as that of  the input (Fig. \ref{Lorfre}(a)) when the coupling strength $\kappa$ is relatively small ($\kappa=0.1$ in Fig. \ref{Lorfre}(b)) or large ($\kappa=2$ in Fig. \ref{Lorfre}(e) and $\kappa=10$ in Fig. \ref{Lorfre}(f)). However, when the coupling strength $\kappa=0.5$, the two output photons can be strongly {\it anti-correlated} in nearly the whole region  except that they are correlated at the origin, see the three parts in Fig. \ref{Lorfre}(c); this has also been observed in cavity opto-mechanical systems \cite{Liao:13}. Moreover, when the detuning is nonzero, for example, $\omega_d=0.1$, the anti-correlation between the  two output photons becomes weak and the maximum value is attained at the origin, see Fig. \ref{Lorfre}(d). Interestingly, in contrast to the joint spectra for Gaussian pulse shapes in Fig. \ref{Gaussian_fre}, anti-correlation is observed in this case of Lorentzian pulse shape (see Fig. \ref{Lorfre}(c)). Such difference means that different pulse shapes give rise to drastically different frequency entanglement.

\begin{remark}
As shown in Fig. \ref{Lorfre}(c), the two output photons can be strongly anti-correlated, which means that there exists a sufficient interaction between the two input photons and the two-level system (or between the photons through the system) when the relative size of the interaction time $1/\kappa$ and the photon lifetime $1/\gamma$ are comparable. On the other hand, when the interaction time is sufficiently small compared to the photon lifetime (Fig. \ref{Lorfre}(f)), the photon-photon interaction is very weak. Finally, when the interaction time is relatively large, the photons cannot ``live" long enough to be absorbed by the two-level system, see Fig. \ref{Lorfre}(b).
\end{remark}
%%%%%%%%%%%%%%%%%%%%%%%
%%%%%%%%%%%%%%%%%%%%%%%

\section{Two-channel case}\label{sec:two-channel}

In this section, we consider the two-level system with two input channels, each containing one photon. The analytic form of the steady-state output field state is presented, in both the time and frequency domains.

\begin{figure}
\begin{center}
\includegraphics[scale=0.25]{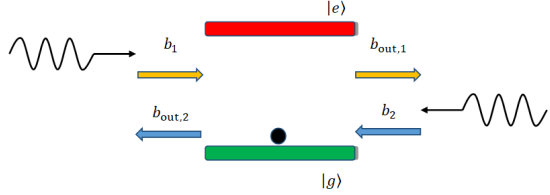}
\caption{(Color online) Two counter-propagating pulsed photons coupled to a two-level system initialized in the ground state.}
\label{scheme}
\end{center}
\end{figure}

The system could be depicted as in Fig. \ref{scheme}. In this scheme, the first output channel $b_{\mathrm{out},1}$ can be regarded as the right-going direction, the second output channel $b_{\mathrm{out},2}$ indicates the left-going direction. The photon $i$ is coupled to the two-level system with the coupling strength $\kappa_i$ ($i=1,2$). Clearly, the input field state is a product state $B_1^\dag(\xi_1)B_2^\dag(\xi_2)|0\rangle$.

\subsection{The output field state in the time domain}\label{two-channel-time}
In this subsection, an analytic form of the output two-photon state is presented in the time domain.

Assume there is no detuning (namely $\omega_d=0$), the system model is
\begin{eqnarray}\label{sys_3a}\begin{aligned}
\dot{\sigma}_-=&-\frac{\kappa_1+\kappa_2}{2}\sigma_-+\sqrt{\kappa_1}\sigma_z(t)b_1(t)+\sqrt{\kappa_2}\sigma_z(t)b_2(t),\\
b_{\mathrm{out},1}(t)=&\sqrt{\kappa_1}\sigma_-(t)+b_1(t),\\
b_{\mathrm{out},2}(t)=&\sqrt{\kappa_2}\sigma_-(t)+b_2(t),\ \ t\geq t_{0}.
\end{aligned}\end{eqnarray}
The initial joint system-field state is
\begin{equation}\label{initial}
\left\vert \Psi (t_{0})\right\rangle =B_{1}^{\dagger }(\xi_{1})B_{2}^{\dagger }(\xi _{2})\left\vert 0g\right\rangle ,
\end{equation}
where $\left\Vert \xi _{1}\right\Vert =\left\Vert \xi _{2}\right\Vert =1$.
At time $t\geq t_0$, the joint system-field state is
\begin{equation}\label{t}
\left\vert \Psi (t)\right\rangle =U(t,t_{0})B_{1}^{\dagger }(\xi_{1})B_{2}^{\dagger }(\xi _{2})\left\vert 0g\right\rangle .
\end{equation}
In analogy with the single-channel case in Section \ref{sec:main-time},  the steady-state output field state is
\begin{equation}\label{infty}\begin{aligned}
\left\vert\Psi_{\mathrm{out}}\right\rangle=\lim_{\begin{subarray}{c}t_0\rightarrow-\infty \\ t\rightarrow\infty\end{subarray}}\left\langle g|\Psi(t)\right\rangle
=\lim_{\begin{subarray}{c}t_0\rightarrow-\infty \\ t\rightarrow\infty\end{subarray}}\left\langle g\right\vert U(t,t_0)B_1^{\dagger}(\xi_1)B_2^{\dagger}(\xi_2)\left\vert0g\right\rangle.
\end{aligned}\end{equation}
Inserting the two-photon basis
\begin{equation*}\begin{aligned}
\bigg\{
&\frac{1}{2}\int dp_1\int dp_2\left\vert1_{1p_1}1_{1p_2}\right\rangle\left\langle1_{1p_1}1_{1p_2}\right\vert,  \ \int dp_1\int dp_2\left\vert1_{1p_1}1_{2p_2}\right\rangle\left\langle1_{1p_1}1_{2p_2}\right\vert,\\
&\frac{1}{2}\int dp_1\int dp_2\left\vert 1_{2p_1}1_{2p_2}\right\rangle\left\langle1_{2p_1}1_{2p_2}\right\vert
\bigg\}
\end{aligned}\end{equation*}
into Eq. (\ref{infty}), the steady-state output field state becomes
\begin{equation}\label{mar17_1}\begin{aligned}
\left\vert\Psi_{\mathrm{out}}\right\rangle
=&\frac{1}{2}\int dp_1\int dp_2\left\vert1_{1p_1}1_{1p_2}\right\rangle
\lim_{\begin{subarray}{c}t_0\rightarrow-\infty \\ t\rightarrow\infty\end{subarray}}\int_{t_0}^tdt_1\int_{t_0}^tdt_2\ \xi_1(t_1)\xi_2(t_2)\\
&\times\sum_{i,j=1}^2\left\langle0g\right\vert b_{\mathrm{out},i}(p_1)b_{\mathrm{out},j}(p_2)b_1^{\dagger}(t_1)b_2^{\dagger}(t_2)\left\vert0g\right\rangle.
\end{aligned}\end{equation}
Thus, in order to derive the analytic form of the steady-state output field state, we need to calculate the following quantities:
\begin{eqnarray}
&&\left\langle 0g\right\vert b_{\mathrm{out},1}(p_{1})b_{\mathrm{out}
,1}(p_{2})b_{1}^{\dagger}(t_{1})b_{2}^{\dagger}(t_{2})\left\vert 0g\right\rangle
,  \label{2_key_a} \\
&&\left\langle 0g\right\vert b_{\mathrm{out},1}(p_{1})b_{\mathrm{out}
,2}(p_{2})b_{1}^{\dagger}(t_{1})b_{2}^{\dagger}(t_{2})\left\vert 0g\right\rangle
,  \label{2_key_b} \\
&&\left\langle 0g\right\vert b_{\mathrm{out},2}(p_{1})b_{\mathrm{out}
,2}(p_{2})b_{1}^{\dagger}(t_{1})b_{2}^{\dagger}(t_{2})\left\vert 0g\right\rangle
,  \label{2_key_c} \\
&&\left\langle 0g\right\vert b_{\mathrm{out},2}(p_{1})b_{\mathrm{out}
,1}(p_{2})b_{1}^{\dagger}(t_{1})b_{2}^{\dagger}(t_{2})\left\vert 0g\right\rangle
.  \label{2_key_d}
\end{eqnarray}
%Notice that
%\begin{equation}\label{mar14_plm}\begin{aligned}
%\left\langle 0g\right\vert b_{\mathrm{out},1}(p_{1})b_{\mathrm{out},2}(p_{2})b_{1}^{\dagger}(t_{1})b_{2}^{\dagger}(t_{2})\left\vert 0g\right\rangle
%=\left\langle 0g\right\vert b_{\mathrm{out},2}(p_{2})b_{\mathrm{out},1}(p_{1})b_{1}^{\dagger}(t_{1})b_{2}^{\dagger}(t_{2})\left\vert 0g\right\rangle.
%\end{aligned}\end{equation}

The calculations of the above quantities are given in Appendix \ref{appendixd}, based on which we can derive the main result of this section, Theorem \ref{thm:2-channel}. The following notations are used in Theorem \ref{thm:2-channel} and Corollary \ref{coro:output1}.
\begin{itemize}
\item Similar to Eq. \eqref{tf_2} in Section \ref{subsec:TLS}, an impulse response function can be defined as 
\begin{equation} \label{tf_3}
g_G(t) \equiv [g_{G_{ij}}(t)] \triangleq\left\{
\begin{array}{cc}
\delta(t)I_{2}-\left[
\begin{array}{c}
\sqrt{\kappa_1} \\
\sqrt{\kappa_2}
\end{array}
\right]e^{-\frac{\kappa_1+\kappa_2}{2}t}\left[
\begin{array}{cc}
\sqrt{\kappa_1} & \sqrt{\kappa_2}
\end{array}
\right], & t\geq 0, \\
0, & t<0.
\end{array}
\right.
\end{equation}
\item The corresponding transfer function is
\begin{equation}\label{transfer2}
G[s] \equiv [G_{mn}[s]]=\frac{1}{s+\frac{\kappa_1+\kappa_2}{2}}
\left[
                  \begin{array}{cc}
                    s-\frac{\kappa_1-\kappa_2}{2} & -\sqrt{\kappa_1\kappa_2} \\
                    -\sqrt{\kappa_1\kappa_2}  & s+\frac{\kappa_1-\kappa_2}{2} \\
                  \end{array}
                \right].
\end{equation}
\item We define  \begin{equation}
\frac{2}{j} \triangleq \left\{\begin{array}{ll}
2, & j=1, \\
1, & j=2.
\end{array} \right.
\end{equation}
\end{itemize}

%The following lemmas simply present the calculations for the quantities given by Eqs. (\ref{2_key_a})-(\ref{2_key_d}), which can be easily verified.

%The steady-state output field state (\ref{mar17_1}) can be re-written as
%\begin{equation}\label{mar31_Psi_out}\begin{aligned}
%|\Psi_{\mathrm{out}}\rangle=&\frac{1}{2}\sum_{i,j=1}^2\int dp_1\int dp_2|1_{ip_1}1_{jp_2}\rangle\int_{-\infty}^\infty dt_1\int_{-\infty}^\infty dt_2\\
%&\times\xi_1(t_1)\xi_2(t_2)
%\langle0g|b_{\mathrm{out},i}(p_1)b_{\mathrm{out},j}(p_2)b_1^\dagger(t_1)b_2^\dagger(t_2)|0g\rangle,
%\end{aligned}\end{equation}
%where $\langle0g|b_{\mathrm{out},i}(p_1)b_{\mathrm{out},j}(p_2)b_1^\dagger(t_1)b_2^\dagger(t_2)|0g\rangle$ is given by Lemma \ref{lemma1}.

%The following is the main result of this section, which follows from Lemma \ref{lemma1} and Eq. \eqref{mar17_1}.

\begin{theorem}\label{thm:2-channel}
The steady-state output field state $|\Psi_{\mathrm{out}}\rangle$ in the time domain is
\begin{equation}\label{June4}\begin{aligned}
|\Psi_{\mathrm{out}}\rangle=\frac{1}{2}\sum_{i,j=1}^2\int_{-\infty}^{\infty}dp_1\int_{-\infty}^{\infty}dp_2\left[\eta_{ij}(p_1,p_2)+\eta_{ij}(p_2,p_1)\right]
b_i^\dag(p_1)b_j^\dag(p_2)|0\rangle,
\end{aligned}\end{equation}
where
\footnotesize
\begin{equation}\label{May20c}
\eta_{ij}(p_1,p_2)=\left\{
\begin{aligned}
&\left[g_{G_{ii}}\ast\xi_i(p_1)\right]\times[g_{G_{\frac{2}{i}j}}\ast\xi_{\frac{2}{i}}(p_2)]
+[g_{G_{ij}}\ast\xi_i(p_2)]\times[g_{G_{\frac{2}{i}i}}\ast\xi_{\frac{2}{i}}(p_1)]\\
&-2\int_{-\infty}^{p_2}dr\left\{\xi_1(r)\left[g_{G_{12}}\ast\xi_2(r)\right]+\xi_2(r)\left[g_{G_{12}}\ast\xi_1(r)\right]\right\}\int_{-\infty}^rd\tau_1\ e^{-\frac{\kappa_1+\kappa_2}{2}(p_2-\tau_1)} \\
&\ \ \times\Big[\kappa_jg_{G_{ij}}\ast\delta(p_1-\tau_1)+\sqrt{\kappa_1\kappa_2}g_{G_{i\frac{2}{j}}}\ast\delta(p_1-\tau_1)\Big],~~p_1\geq p_2, \\
%&-2\int_{-\infty}^{p_2}dr\ e^{-\frac{\kappa_1+\kappa_2}{2}(p_2-r)}\xi_{\frac{2}{i}}(r)\left[g_{G_{\frac{2}{i}i}}\ast\xi_i(r)\right]\int_{-\infty}^rd\tau_1\ e^{-\frac{\kappa_1+\kappa_2}{2}(r-\tau_1)}\\
%&\ \ \ \times
%\Big[\kappa_jg_{G_{ij}}\ast\delta(p_1-\tau_1)+\sqrt{\kappa_1\kappa_2}g_{G_{i\frac{2}{j}}}\ast\delta(p_1-\tau_1)\Big],~~p_1\geq p_2, \\
&0,~~p_1<p_2,
\end{aligned}\right.
\end{equation}
\normalsize
for $i,j=1,2$. In particular, if $\kappa_1=\kappa_2=\kappa$, $\xi_1=\xi_2=\xi$,  in other words, the two-level system is equally coupled to two indistinguishable input photons, we have $\eta_{11}(p_1,p_2)=\eta_{22}(p_1,p_2)$, $\eta_{12}(p_1,p_2)=\eta_{21}(p_1,p_2)$, where
\begin{equation}\label{May20b}
\eta_{11}(p_1,p_2)
=\left[g_{G_{11}}\ast\xi(p_1)\right]\times \left[g_{G_{12}}\ast\xi(p_2)\right]
+\left[g_{G_{11}}\ast\xi(p_2)\right]\times \left[g_{G_{12}}\ast\xi(p_1)\right]+\chi(p_1,p_2),
\end{equation}
\begin{equation}\label{May20}
\eta_{12}(p_1,p_2)
=\left[g_{G_{11}}\ast\xi(p_1)\right]\times \left[g_{G_{11}}\ast\xi(p_2)\right]
+\left[g_{G_{12}}\ast\xi(p_2)\right]\times \left[g_{G_{12}}\ast\xi(p_1)\right]+\chi(p_1,p_2),
\end{equation}
with
\begin{equation}\label{May20a}
\chi(p_1,p_2)=\left\{\begin{aligned}
                       &4\kappa\ e^{-\kappa (p_1+ p_2)}\int_{-\infty}^{p_2}dr\ e^{2\kappa r}\xi(r)\left[g_{G_{12}}\ast\xi(r)\right],~~p_1\geq p_2, \\
                       &0,~~p_1<p_2.
                     \end{aligned}\right.
\end{equation}
In this case, the resulting steady-state output field state is
\begin{equation}\begin{aligned}
|\Psi_{\mathrm{out}}\rangle=&\frac{1}{2}\int_{-\infty}^{\infty}dp_1\int_{-\infty}^{\infty}dp_2\left[\eta_{11}(p_1,p_2)+\eta_{11}(p_2,p_1)\right]b_1^\dag(p_1)b_1^\dag(p_2)|0\rangle \\
&+\int_{-\infty}^{\infty}dp_1\int_{-\infty}^{\infty}dp_2\left[\eta_{12}(p_1,p_2)+\eta_{12}(p_2,p_1)\right]b_1^\dag(p_1)b_2^\dag(p_2)|0\rangle \\
&+\frac{1}{2}\int_{-\infty}^{\infty}dp_1\int_{-\infty}^{\infty}dp_2\left[\eta_{11}(p_1,p_2)+\eta_{11}(p_2,p_1)\right]b_2^\dag(p_1)b_2^\dag(p_2)|0\rangle.
\end{aligned}\end{equation}
\end{theorem}

\begin{remark}
The first two terms in Eqs. (\ref{May20c})-(\ref{May20}) represent the single-photon scattering processes in the two-photon scattering scheme, the third term is the temporal correlation between the output photons induced by the two-level system \cite{Nysteen15,Roulet16}, which are called the background fluorescence in \cite{Shen:07}.
\end{remark}

\subsection{The output field state in the frequency domain}\label{two-channel-fre}

Similarly,  as in the one-channel case discussed before, by applying the Fourier transform to the time variables $p_1$, $p_2$ in Eq. (\ref{June4}), the steady-state output field state in the frequency domain can be obtained.
%\begin{equation}\label{June4a}\begin{aligned}
%&|\Psi_{\mathrm{out}}\rangle\\
%=&\frac{1}{2}\int_{-\infty}^{\infty}d\omega_1\int_{-\infty}^{\infty}d\omega_2|1_{1_{\omega_1}}1_{1_{\omega_2}}\rangle
%\int_{-\infty}^{\infty}d\mu_1\int_{-\infty}^{\infty}d\mu_2\xi_1[\mu_1]\xi_2[\mu_2]\\
%&\times\langle0g|b_{\mathrm{out},1}[\omega_1]b_{\mathrm{out},1}[\omega_2]b_1^\dagger[\mu_1]b_2^\dagger[\mu_2]|0g\rangle\\
%&+\int_{-\infty}^{\infty}d\omega_1\int_{-\infty}^{\infty}d\omega_2|1_{1_{\omega_1}}1_{2_{\omega_2}}\rangle
%\int_{-\infty}^{\infty}d\mu_1\int_{-\infty}^{\infty}d\mu_2\xi_1[\mu_1]\xi_2[\mu_2]\\
%&\times\langle0g|b_{\mathrm{out},1}[\omega_1]b_{\mathrm{out},2}[\omega_2]b_1^\dagger[\mu_1]b_2^\dagger[\mu_2]|0g\rangle\\
%&+\frac{1}{2}\int_{-\infty}^{\infty}d\omega_1\int_{-\infty}^{\infty}d\omega_2|1_{2_{\omega_1}}1_{2_{\omega_2}}\rangle
%\int_{-\infty}^{\infty}d\mu_1\int_{-\infty}^{\infty}d\mu_2\xi_1[\mu_1]\xi_2[\mu_2]\\
%&\times\langle0g|b_{\mathrm{out},2}[\omega_1]b_{\mathrm{out},2}[\omega_2]b_1^\dagger[\mu_1]b_2^\dagger[\mu_2]|0g\rangle.
%\end{aligned}\end{equation}

\begin{corollary}\label{coro:output1}
The steady-state output field state $|\Psi_{\mathrm{out}}\rangle$ is
\begin{equation}\label{June4b}\begin{aligned}
\left\vert\Psi_{\mathrm{out}}\right\rangle
=&\; \frac{1}{2}\int_{-\infty}^{\infty}d\omega_1\int_{-\infty}^{\infty}d\omega_2\
T_{11}[\omega_1,\omega_2]b_1^\dagger[\omega_1]b_1^\dagger[\omega_2]|0\rangle\\
&+\int_{-\infty}^{\infty}d\omega_1\int_{-\infty}^{\infty}d\omega_2\
T_{12}[\omega_1,\omega_2]b_1^\dagger[\omega_1]b_2^\dagger[\omega_2]|0\rangle\\
&+\frac{1}{2}\int_{-\infty}^{\infty}d\omega_1\int_{-\infty}^{\infty}d\omega_2\
T_{22}[\omega_1,\omega_2]b_2^\dagger[\omega_1]b_2^\dagger[\omega_2]|0\rangle,
\end{aligned}\end{equation}
where
\begin{equation}\label{June4c1}\begin{aligned}
&T_{11}[\omega_1,\omega_2]\\
=&\; G_{11}[\mathrm{i}\omega_1]G_{12}[\mathrm{i}\omega_2]\xi_1[\omega_1]\xi_2[\omega_2]+G_{11}[\mathrm{i}\omega_2]G_{12}[\mathrm{i}\omega_1]\xi_1[\omega_2]\xi_2[\omega_1] \\
&+\frac{\sqrt{\kappa_1\kappa_2}}{\pi\kappa_1^2}\int_{-\infty}^{\infty}d\mu_1\
\xi_1[\mu_1]\xi_2[\omega_1+\omega_2-\mu_1]g(\omega_1,\omega_2,\mu_1,\omega_1+\omega_2-\mu_1),
\end{aligned}\end{equation}
\begin{equation}\label{June4c2}\begin{aligned}
&T_{12}[\omega_1,\omega_2]\\
=&\; G_{11}[\mathrm{i}\omega_1]G_{22}[\mathrm{i}\omega_2]\xi_1[\omega_1]\xi_2[\omega_2]+G_{12}[\mathrm{i}\omega_1]G_{12}[\mathrm{i}\omega_2]\xi_1[\omega_2]\xi_2[\omega_1] \\
&+\frac{\kappa_2}{\pi\kappa_1^2}\int_{-\infty}^{\infty}d\mu_1\
\xi_1[\mu_1]\xi_2[\omega_1+\omega_2-\mu_1] g(\omega_1,\omega_2,\mu_1,\omega_1+\omega_2-\mu_1),
\end{aligned}\end{equation}
\begin{equation}\label{June4c3}\begin{aligned}
&T_{22}[\omega_1,\omega_2]\\
=& \; G_{12}[\mathrm{i}\omega_1]G_{22}[\mathrm{i}\omega_2]\xi_1[\omega_1]\xi_2[\omega_2]+G_{12}[\mathrm{i}\omega_2]G_{22}[\mathrm{i}\omega_1]\xi_1[\omega_2]\xi_2[\omega_1] \\
&+\frac{\kappa_2\sqrt{\kappa_1\kappa_2}}{\pi\kappa_1^3}\int_{-\infty}^{\infty}d\mu_1\
\xi_1[\mu_1]\xi_2[\omega_1+\omega_2-\mu_1] g(\omega_1,\omega_2,\mu_1,\omega_1+\omega_2-\mu_1),
\end{aligned}\end{equation}
and
\begin{equation}\label{June4d1}
g(\omega_1,\omega_2,\mu_1,\mu_2)=(G_{11}[\mathrm{i}\omega_1]-1)(G_{11}[\mathrm{i}\omega_2]-1)(G_{11}[\mathrm{i}\mu_1]+G_{11}[\mathrm{i}\mu_2]-2),
\end{equation}
with $G_{mn}[s]$ given by Eq. (\ref{transfer2}).
\end{corollary}

\begin{remark} \label{rem:may27_1}
If $\kappa_1=\kappa_2=\kappa$, then the steady-state output field state $|\Psi_{\mathrm{out}}\rangle$ in (\ref{June4b}) has the same form of the postscattering state in \cite[Eq. (28)]{Nysteen15}. Moreover, the four-wave mixing processes, i.e., the terms containing $g(\omega_1,\omega_2,\mu_1,\omega_1+\omega_2-\mu_1)$ in Eqs. (\ref{June4c1})-(\ref{June4c3}), are related to the nonlinear frequency entanglement of two-photon scattering. The output photons with frequencies $\omega_1$ and $\omega_2$ can be generated by any pair of incident photons with frequencies $\mu_1$, $\mu_2$ satisfying $\omega_1+\omega_2=\mu_1+\mu_2$. That is, the sum of the energies of the two input  photons is conserved. In addition, the functions $T_{ij}[\omega_1,\omega_2]$ ($i,j=1,2$) should be symmetric, i.e., $T_{ij}[\omega_1,\omega_2] = T_{ij}[\omega_2,\omega_1] $. However, this is hard to see from the forms given above, due to their complex form. Nevertheless, the numerical simulations presented in the next section clearly reveal the symmetry required.
\end{remark}

%Let $\kappa_1+\kappa_2=\kappa$, $\omega_d=0$, the function $g(\omega_1,\omega_2,\mu_1,\mu_2)$ in Eq. (\ref{June4d1}) reduces to that in Eq. (\ref{May23d}).

\subsection{Numerical example}

\emph{Example} 3. In the aid of the two analytic forms of the two-photon output field state derived above, we are able to compute various physical quantities. As demonstration, in this example we compute the probabilities of finding the two photons in different directions. These probabilities are visualized in both the time and frequency domains.   The pulse shapes of the two input photons are given respectively by $\xi_i(t)=-\sqrt{\gamma_i}e^{\frac{\gamma_i}{2}t}(1-u(t))$, ($i=1,2$), where $u(t)$ is  the Heaviside function defined in Eq. (\ref{Hea}). For simplicity, we assume that the input two photons have the same pulse shapes, i.e., $\gamma_1=\gamma_2=\gamma$, and are equally coupled to the two-level system, $\kappa_1=\kappa_2=\kappa$.

\begin{figure}
\begin{center}
\includegraphics[scale=0.4]{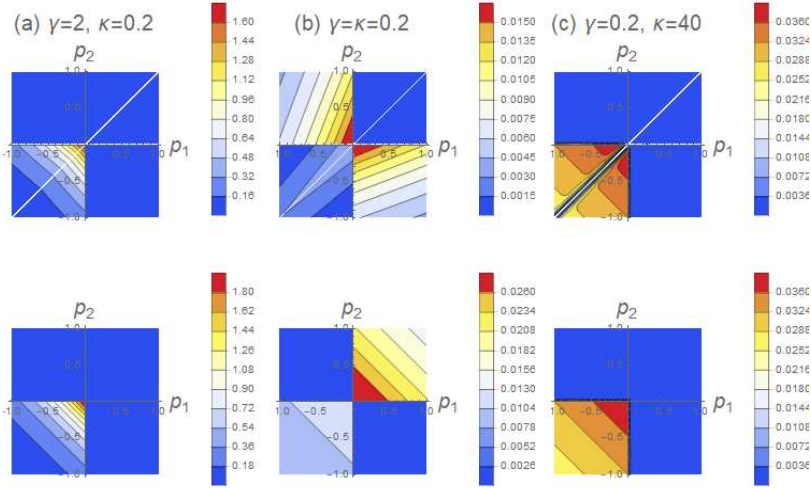}
\caption{(Color online) The time distributions $|\eta_{12}(p_1,p_2)+\eta_{12}(p_2,p_1)|^2$ of the  two output photons scattering in different directions. The first row corresponds to the probabilities of the two photons being scattered into different channels. For comparison, the second row shows the linear single-photon scattering processes with the nonlinear term $\chi(p_1,p_2)$ in Eq. (\ref{May20}) being removed.}
\label{fig_May20}
\end{center}
\end{figure}

Firstly, we focus on the time distribution of the  two output photons scattering in different directions, for which $|\eta_{12}(p_1,p_2)+\eta_{12}(p_2,p_1)|^2$ in Eq. (\ref{May20}) is  plotted in Fig. \ref{fig_May20}. As the two-level system can only absorb a single photon each time or spontaneously emit a single photon, the time distributions vanish for $p_1, p_2>0$, i.e., $\eta_{12}(p_1,p_2)+\eta_{12}(p_2,p_1)\equiv0$ for $p_1,p_2>0$ as can be seen in the first row of Fig. \ref{fig_May20}. Actually, this can be verified by Theorem \ref{thm:2-channel} directly.  When $\gamma\gg\kappa$, the two photons do not live long enough for sufficient interaction with the two-level system, thus the time distribution is barely modified by  the two-level system, as shown in Fig. \ref{fig_May20}(a). This is consistent with the case discussed in \cite[Fig. 7]{Roulet16}.  When the bandwidth $\gamma$ is comparable to coupling $\kappa$, the presence of two valleys ($|\eta_{12}(p_1,p_2)+\eta_{12}(p_2,p_1)|^2 \approx0$) in the region $p_1, p_2\leq 0$ in Fig. \ref{fig_May20}(b) demonstrates the signature of the nonlinearity induced by the two-level system; in other words, it is impossible to observe the two photons in different output channels. Such nonlinearity cannot be found in the linear single-photon scattering processes. When $\gamma\ll\kappa$, the lifetime of the two-level system is too short, or in other words, the energies of  two input photons are too spread out for efficient excitation, the two-level system acts as a fully reflecting mirror, and the strongest nonlinearity can be attained in the two valleys close to the diagonal $p_1=p_2$ as shown in Fig. \ref{fig_May20}(c); this is  also consistent with \cite[Fig. 7]{Roulet16}.

\begin{figure}[htp!]
\begin{center}
\includegraphics[scale=0.4]{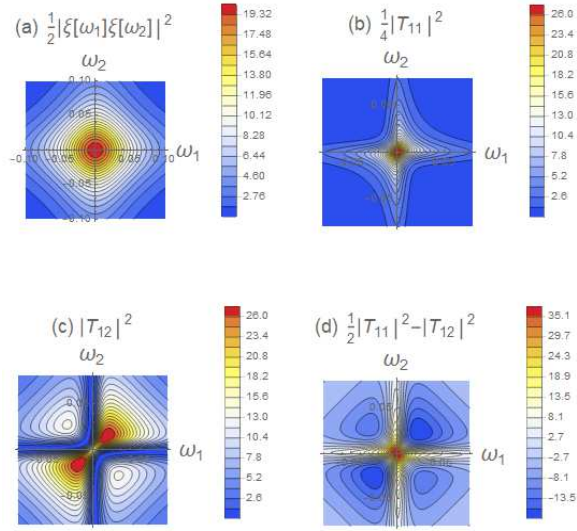}
\caption{(Color online) $\gamma=0.1$, $\kappa=0.01$.}
\label{two_channel_1}
\end{center}
\end{figure}

\begin{figure}[htp!]
\begin{center}
\includegraphics[scale=0.4]{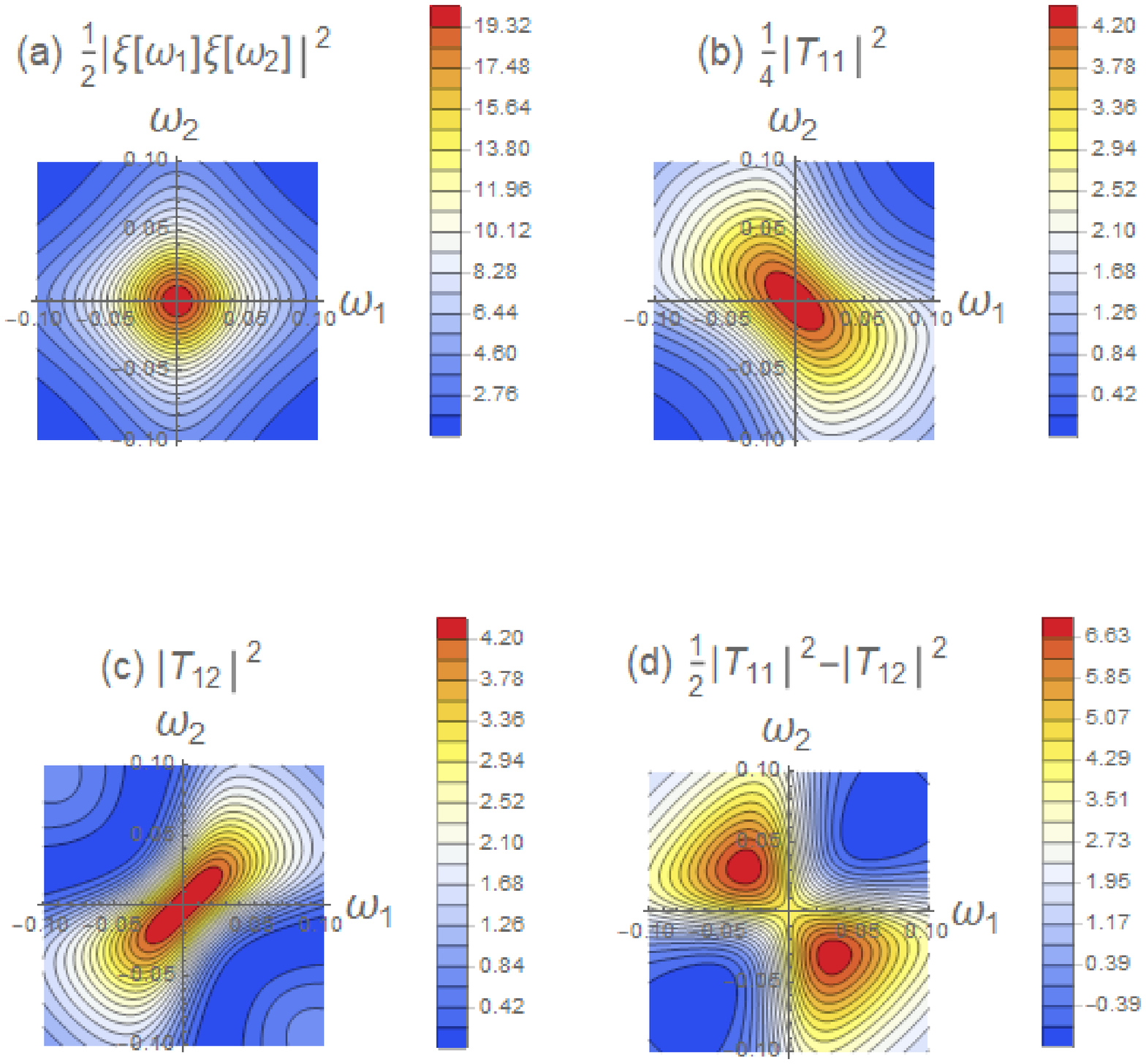}
\caption{(Color online) $\gamma=0.1$, $\kappa=0.1$.}
\label{two_channel_2}
\end{center}
\end{figure}

\begin{figure}[htp!]
\begin{center}
\includegraphics[scale=0.4]{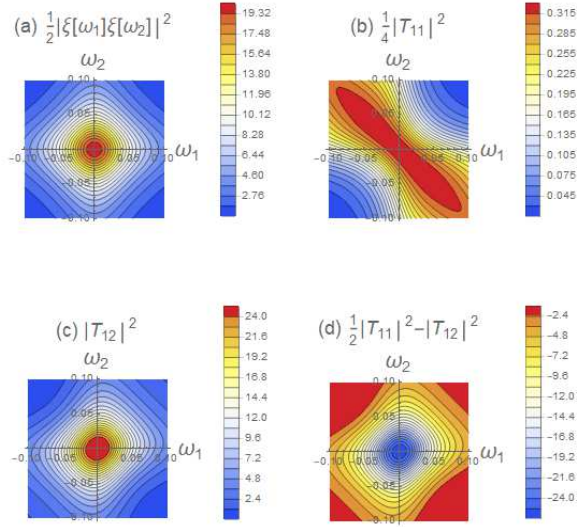}
\caption{(Color online) $\gamma=0.1$, $\kappa=0.5$.}
\label{two_channel_3}
\end{center}
\end{figure}

In the following,  we fix $\gamma=0.1$  and study the output two-photon joint spectra for different couplings $\kappa$. In Fig. \ref{two_channel_1},  the input two-photon joint spectrum is shown in Fig. \ref{two_channel_1}(a), the joint spectra for the  two output photons in either the first or the second channel are given in Fig. \ref{two_channel_1}(b), the joint spectra for each channel containing one output photons are provided in Fig. \ref{two_channel_1} (c), and Fig. \ref{two_channel_1}(d) shows the difference of the joint spectra between Fig. \ref{two_channel_1}(b) and (c). The same settings hold in Figs. \ref{two_channel_2}-\ref{two_channel_3}. We have the following observations.

\begin{description}
  \item[(i)] In Fig. \ref{two_channel_1}, when the coupling strength is very small ($\kappa=0.01$) compared with $\gamma$, it can be seen that the values of two-photon spectra are rather small in most regions away from the origin; see Fig. \ref{two_channel_1}(b). On the other hand, when each channel contains exactly one output photon, the two photons become correlated, see Fig. \ref{two_channel_1}(c). In Fig. \ref{two_channel_1}(d), the two output photons exhibit the HOM bunching effect \cite{HOM:87} {\it only} in the frequency region $(\omega_1,\omega_2)\approx(0,0)$, and they are mostly in the different channels in the other frequency regions. Similar observations can be found in \cite[Fig. 5]{Nysteen15}.

  \item[(ii)] In Fig. \ref{two_channel_2},  $\kappa=\gamma=0.1$. In this case, if the  two output photons are in the same channel (Fig. \ref{two_channel_2}(b)), they are strongly anti-correlated. This demonstrates the four-wave mixing nonlinear effect; see Remark \ref{rem:may27_1}. In contrast, if each channel contains exactly one output photon (Fig. \ref{two_channel_2}(c)), the  two output photons are strongly correlated. This scenario is consistent with  \cite[Fig. 5]{Nysteen15} and \cite[Fig. 6]{Roulet16}. Finally, as can be seen in Fig. \ref{two_channel_2}(d), along the line $\omega_1+\omega_2=0$ the HOM bunching effect is prominent.

  \item[(iii)] In Fig. \ref{two_channel_3}, we choose $\kappa=0.5$. As pointed out by \cite[Fig. 6]{Roulet16}, the two-level system was found to be linear and shape preserving when $\kappa\gg\gamma$. Thus in this case, the  two output photons are mainly reflected (Fig. \ref{two_channel_3}(d)) and the joint spectra are similar to that of the input, cf. Fig. \ref{two_channel_3}(a) and (c). On the other hand, if one of them is indeed transmitted, the  two output photons are strongly anti-correlated (Fig. \ref{two_channel_3}(b)), which is similar to the $(11,22)$ case in \cite[Fig. 5]{Nysteen15}. This is also consistent with the result of two-photon transport in a Kerr nonlinear cavity \cite{Liao:10}.
\end{description}

%\begin{remark}
%\color{red}The output two photons are strongly correlated/anti-correlated and the photon-photon interaction induced by the two-level atom is maximized when the coupling strength is equal to the FWHM of the input photons in Fig. \ref{two_channel_2}. This can be understood by analyzing the average coincidence at the output of the two-level atom as a function of the normalized bandwidth $\Omega/\gamma$ in \cite[Fig. 4]{Roulet16}. The entanglement correlation of the two photons induced by the nonlinearity of the two-level atom can only be observed clearly when $\Omega\approx\gamma$, which is equivalent to the average coincidence is less than $1/2$.
%\end{remark}

\section{Conclusion}\label{sec:con}

In this paper, the response of a two-level system to two-photon inputs has been investigated. The output two-photon states have been explicitly derived in both the time and frequency domains when the two photons are either in the same channel or    counter-propagating  along different directions. For both cases, simulation results have demonstrated rich and interesting properties of the output two-photon states. Future research includes the applications of the theoretical results in the field of quantum communication and quantum computing.

\appendix

\section{Proof of Theorem \ref{thm:output}}\label{appendixa}
In this appendix, we firstly prove Lemma \ref{lem:out4} which presents a form of the output two-photon pulse shape;  after that, we simplify it to get the final expression as given in Theorem \ref{thm:output}.

\begin{lemma}\label{lem:out4}
The steady-state output field state $\left\vert\Psi _{\mathrm{out}}\right\rangle$ in Eq. (\ref{eq:feb20_1}) can be calculated as
\begin{equation}\label{eq:out_3}\begin{aligned}
\left\vert\Psi_{\mathrm{out}}\right\rangle=\frac{1}{2\sqrt{N_{2}}}\int_{-\infty }^{\infty }dp_{1}\int_{-\infty }^{\infty }dp_{2}\
\eta(p_{1},p_{2})b^{\dag }(p_{1})b^{\dag }(p_{2})\left\vert 0\right\rangle,
\end{aligned}\end{equation}
where
\begin{equation}\label{eta_5}\begin{aligned}
&\eta(p_1,p_2)\\
\triangleq&\xi_1(p_2)\xi_2(p_1)+\xi_1(p_1)\xi_2(p_2)\\
&-\kappa\int_{-\infty}^{p_1}d\tau\ e^{-\left(\frac{\kappa}{2}+\mathrm{i}\omega_d\right)(p_1-\tau)}[\xi_1(p_2)\xi_2(\tau)+\xi_2(p_2)\xi_1(\tau)]\\
&-\kappa\int_{-\infty}^{p_2}d\tau\ e^{-\left(\frac{\kappa}{2}+\mathrm{i}\omega_d\right)(p_2-\tau)}[\xi_1(p_1)\xi_2(\tau)+\xi_2(p_1)\xi_1(\tau)]\\
&+\kappa^2\int_{-\infty}^{p_2}d\tau\int_{-\infty}^{p_1}dr\ e^{-\left(\frac{\kappa}{2}+\mathrm{i}\omega_d\right)(p_1+p_2-\tau-r)}
[\xi_1(r)\xi_2(\tau)+\xi_2(r)\xi_1(\tau)]\\
&+\kappa^2\bigg[\int_{-\infty}^{p_1}d\tau\ e^{-\left(\frac{\kappa}{2}+\mathrm{i}\omega_d\right)(p_1-\tau)}\int_{-\infty}^{\tau}d\tau_1\ \delta(\tau_1-p_2)\\
&\ \ \ \ \ \ \ \ +\int_{-\infty}^{p_2}d\tau\ e^{-\left(\frac{\kappa}{2}+\mathrm{i}\omega_d\right)(p_2-\tau)}\int_{-\infty}^{\tau}d\tau_1\ \delta(\tau_1-p_1)\bigg]\\
&\ \ \ \times e^{-\left(\frac{\kappa}{2}-\mathrm{i}\omega_d\right)(\tau-\tau_1)}\int_{-\infty}^{\tau}d\tau_2\ e^{-\left(\frac{\kappa}{2}+\mathrm{i}\omega_d\right)(\tau -\tau_2)}
[\xi _{1}(\tau_{2})\xi _{2}(\tau )+\xi _{2}(\tau _{2})\xi _{1}(\tau )]\\
&-\kappa^3\int_{-\infty}^{p_1}dr\int_{-\infty}^{p_2}d\tau\ e^{-\left(\frac{\kappa}{2}+\mathrm{i}\omega_d\right)(p_1+p_2-\tau-r)}\\
&\ \ \ \times\left\{\int_{-\infty}^rd\tau_1\ e^{-\left(\frac{\kappa}{2}-\mathrm{i}\omega_d\right)(r-\tau_1)}\delta(\tau_1-\tau)\int_{-\infty}^rd\tau_2\right.\\
&\ \ \ \ \ \ \ \ \ \ \ \times e^{-\left(\frac{\kappa}{2}+\mathrm{i}\omega_d\right)(r-\tau_2)}[\xi_2(r)\xi_1(\tau_2)+\xi_1(r)\xi_2(\tau_2)]\\
&\ \ \ \ \ \ \ \ +\int_{-\infty}^{\tau}d\tau_1\ e^{-\left(\frac{\kappa}{2}-\mathrm{i}\omega_d\right)(\tau-\tau_1)}\delta(\tau_1-r)\int_{-\infty}^{\tau}d\tau_2\\
&\ \ \ \ \ \ \ \ \ \ \ \times e^{-\left(\frac{\kappa}{2}+\mathrm{i}\omega_d\right)(\tau-\tau_2)}[\xi_2(\tau)\xi_1(\tau_2)+\xi_1(\tau)\xi_2(\tau_2)]\bigg\}.
\end{aligned}\end{equation}
\end{lemma}

{\it Proof.} Firstly, by Eq. (\ref{eq:b_out2}) and $\left\langle 0g\right\vert U(t,t_{0})=\left\langle 0g\right\vert$, for $t\geq\max\{p_{1},p_{2}\}\geq t_{0}$ (this can always be guaranteed because we are interested in the steady state case $t_{0}\rightarrow-\infty $ and $t\rightarrow \infty$), we have
\begin{equation}\label{eq:feb22_2}\begin{aligned}
\left\langle1_{p_1}1_{p_2}g\right\vert U(t,t_0)b^{\dag}(t_1)b^{\dag}(t_2)\left\vert 0g\right\rangle
=\left\langle 0g\right\vert b_{\mathrm{out}}(p_1)b_{\mathrm{out}}(p_2)b^{\dag}(t_1)b^{\dag}(t_2)\left\vert 0g\right\rangle.
\end{aligned}\end{equation}
Secondly, by substituting Eq. (\ref{eq:feb22_2}) into Eq. (\ref{eq:feb22_1}),  we get
\begin{equation}\label{eq:feb23_temp1}\begin{aligned}
\left\vert\Psi_{\mathrm{out}}\right\rangle
=&\frac{1}{2\sqrt{N_2}}\int_{-\infty}^{\infty}dp_1\int_{-\infty}^{\infty}dp_2\left\vert1_{p_1}1_{p_2}\right\rangle
\lim_{\begin{subarray}{c}t_0\rightarrow-\infty \\ t\rightarrow\infty\end{subarray}}\int_{t_0}^{t}dt_1\xi_1(t_1)\int_{t_0}^{t}dt_2\xi_2(t_2)\\
&\times\left\langle0g\right\vert b_{\mathrm{out}}(p_1)b_{\mathrm{out}}(p_2)b^{\dag}(t_1)b^{\dag}(t_2)\left\vert 0g\right\rangle.
\end{aligned}\end{equation}
Finally, by Eq. (\ref{output}) we have
\begin{equation}\label{temp_11}\begin{aligned}
&\left\langle 0g|b_{\mathrm{out}}(p_{1})b_{\mathrm{out}}(p_{2})b^{\dag}(t_{1})b^{\dag }(t_{2})|0g\right\rangle\\
=&\sqrt{\kappa }\left\langle 0g|\sigma _{-}(p_{1})b_{\mathrm{out}}(p_{2})b^{\dag }(t_{1})b^{\dag }(t_{2})|0g\right\rangle
+\left\langle0g|b(p_{1})b_{\mathrm{out}}(p_{2})b^{\dag }(t_{1})b^{\dag}(t_{2})|0g\right\rangle .
\end{aligned}\end{equation}
Substituting Eq. (\ref{temp_11}) into Eq. (\ref{eq:feb23_temp1}) yields the steady-state output field state $\left\vert \Psi_{\mathrm{out}}\right\rangle$,
\begin{equation}\label{eq:feb23_output}
\left\vert\Psi _{\mathrm{out}}\right\rangle \\
= \frac{1}{2\sqrt{N_{2}}}\int_{-\infty }^{\infty }dp_{1}\int_{-\infty}^{\infty }dp_{2}\ \Gamma(p_1,p_2)  \left\vert 1_{p_{1}}1_{p_{2}}\right\rangle,
\end{equation}
where
\begin{equation}\label{eq.Gamma_p1p2}\begin{aligned}
\Gamma(p_1,p_2)&=\lim_{\begin{subarray}{c}t_0\rightarrow-\infty \\ t\rightarrow\infty\end{subarray}}\int_{t_{0}}^{t}dt_{1}\
\xi _{1}(t_{1})\int_{t_{0}}^{t}dt_{2}\ \xi _{2}(t_{2})\times \\
&\Big[\sqrt{\kappa }\left\langle 0g|\sigma _{-}(p_{1})b_{\mathrm{out}}(p_{2})b^{\dag }(t_{1})b^{\dag }(t_{2})|0g\right\rangle
+\left\langle0g|b(p_{1})b_{\mathrm{out}}(p_{2})b^{\dag }(t_{1})b^{\dag}(t_{2})|0g\right\rangle\Big].
\end{aligned}\end{equation}
Thus, to derive the steady-state output field state $\left\vert \Psi _{\mathrm{out}}\right\rangle$, we have to calculate the two terms on the right-hand side of Eq. (\ref{eq.Gamma_p1p2}), namely, $\left\langle 0g|\sigma_-(p_1)b_{\mathrm{out}}(p_2)b^{\dag}(t_1)b^{\dag}(t_2)|0g\right\rangle$, and $\left\langle0g|b(p_{1})b_{\mathrm{out}}(p_{2})b^{\dag}(t_1)b^{\dag}(t_2)|0g\right\rangle$. Due to page limit, these calculations are omitted, and the final expressions are  given below.
\footnotesize
\begin{equation}\label{eq:out_2a}\begin{aligned}
&\left\langle0g|\sigma_-(p_1)b_{\mathrm{out}}(p_2)b^{\dag}(t_1)b^{\dag}(t_2)|0g\right\rangle\\
%=&-2\kappa^{3/2}e^{-\left(\frac{\kappa }{2}+\mathrm{i}\omega_d\right)(p_{1}-t_{0})}\int_{t_{0}}^{p_{2}}dr\ e^{-\left(\frac{\kappa }{2}+\mathrm{i}\omega_d\right)(p_{2}-r)}e^{-\left( \frac{\kappa}{2}-\mathrm{i}\omega_d\right)(r-t_{0})}\\
%&\ \ \ \times\int_{t_0}^{r}d\tau\ e^{-\left(\frac{\kappa }{2}+\mathrm{i}\omega_d\right)(r-\tau)}[\delta(r-t_{1})\delta (\tau -t_{2})+\delta (r-t_{2})\delta(\tau -t_1)]\\
=&-\sqrt{\kappa }\int_{t_{0}}^{p_{1}}dr\ e^{-\left(\frac{\kappa }{2}+\mathrm{i}\omega_d \right) (p_{1}-r)}\left[ \delta (p_{2}-t_{1})\delta(r-t_{2})+\delta (r-t_{1})\delta (p_{2}-t_{2})\right]\\
&+\kappa ^{3/2}\int_{t_{0}}^{p_{1}}dr\ \int_{t_{0}}^{p_{2}}dn\ e^{-\left(\frac{\kappa }{2}+\mathrm{i}\omega_d \right) (p_{1}+p_{2}-r-n)}[\delta(r-t_{1})\delta (n-t_{2})+\delta (n-t_{1})\delta (r-t_{2})]\\
&-2\kappa^{5/2}\int_{t_{0}}^{p_{1}}dr\ \int_{t_{0}}^{p_{2}}dn\ e^{-\left(\frac{\kappa }{2}+\mathrm{i}\omega_d\right)(p_{1}+p_{2}-r-n)}\int_{t_{0}}^{n}d\tau _{1}\ e^{-\left( \frac{\kappa }{2}-\mathrm{i}\omega_d \right) (n-\tau _{1})}\delta(\tau_1-r)\\
&\ \ \ \times\int_{t_{0}}^{n}d\tau _{2}\ e^{-\left( \frac{\kappa }{2}+\mathrm{i}\omega_d \right) (n-\tau _{2})}[\delta (\tau _{2}-t_{1})\delta(n-t_2)+\delta(n-t_1)\delta (\tau_2-t_2)],
\end{aligned}\end{equation}
\normalsize

%Notice that, due to the presence of the factor $e^{-\left(\frac{\kappa}{2}+\mathrm{i}\omega_d\right)(p_1-t_0)}$, the first term in Eq. (\ref{eq:out_2a}) vanishes in the limit $t_0\rightarrow-\infty$.
%We have
and
\footnotesize
\begin{equation}\label{eq:out_2b}\begin{aligned}
&\left\langle0g|b(p_1)b_{\mathrm{out}}(p_2)b^{\dag}(t_1)b^{\dag}(t_2)|0g\right\rangle\\
=&\delta(p_2-t_1)\delta (p_1-t_2)+\delta(p_1-t_1)\delta(p_2-t_2)\\
&-\kappa\int_{t_0}^{p_2}dr\ e^{-\left(\frac{\kappa}{2}+\mathrm{i}\omega_d\right)(p_2-r)}[\delta(p_1-t_1)\delta(r-t_2)+\delta(r-t_1)\delta(p_1-t_2)]\\
&+2\kappa^2\int_{t_0}^{p_2}dr\ e^{-\left(\frac{\kappa}{2}+\mathrm{i}\omega_d\right)(p_2-r)}\int_{t_0}^rd\tau_1\ e^{-\left(\frac{\kappa}{2}-\mathrm{i}\omega_d \right)(r-\tau_1)}\delta(\tau_1-p_1)\\
&\ \ \ \times\int_{t_0}^rd\tau_2\ e^{-\left(\frac{\kappa}{2}+\mathrm{i}\omega_d\right)(r-\tau_2)}[\delta (\tau_2-t_1)\delta(r-t_2)+\delta(r-t_1)\delta (\tau_2-t_2)].
\end{aligned}\end{equation}
\normalsize
Inserting Eqs. (\ref{eq:out_2a})-(\ref{eq:out_2b}) into Eq. (\ref{eq.Gamma_p1p2}) yields

\begin{equation}\label{eta_3}\begin{aligned}
&\Gamma(p_{1},p_{2})\\
=&\; \xi_{1}(p_{1})\xi_{2}(p_{2})+\xi_{1}(p_{2})\xi_{2}(p_{1})\\
&-\kappa\int_{-\infty}^{p_2}d\tau\ e^{-\left(\frac{\kappa }{2}+\mathrm{i}\omega_d \right)(p_{2}-\tau)}[\xi_{1}(p_{1})\xi_{2}(\tau)+\xi_{2}(p_{1})\xi_{1}(\tau)]\\
&-\kappa\int_{-\infty}^{p_1}d\tau\ e^{-\left(\frac{\kappa }{2}+\mathrm{i}\omega_d \right)(p_{1}-\tau)}[\xi_{1}(p_{2})\xi_{2}(\tau)+\xi_{2}(p_{2})\xi_{1}(\tau)]\\
&+\kappa^2\int_{-\infty}^{p_1}d\tau\int_{-\infty}^{p_2}dr\ e^{-\left(\frac{\kappa }{2}+\mathrm{i}\omega_d\right) (p_1+p_2-\tau-r)}
[\xi_{1}(r)\xi_{2}(\tau )+\xi_{2}(r)\xi_{1}(\tau)]\\
&+2\kappa^2\int_{-\infty }^{p_{2}}d\tau\ e^{-\left( \frac{\kappa}{2}+\mathrm{i}\omega_d \right) (p_{2}-\tau )}\int_{-\infty}^{\tau}d\tau_1\ e^{-\left(\frac{\kappa }{2}-\mathrm{i}\omega_d \right)(\tau-\tau_1)}\\
&\ \ \ \times\delta(\tau_1-p_1)\int_{-\infty}^{\tau}d\tau_2\ e^{-\left(\frac{\kappa}{2}+i\omega_d \right) (\tau-\tau_2)}
[\xi _{1}(\tau _{2})\xi_{2}(\tau )+\xi_{2}(\tau_{2})\xi _{1}(\tau )]\\
&-2\kappa ^3\int_{-\infty }^{p_{1}}d\tau\int_{-\infty }^{p_2}dr\ e^{-\left(\frac{\kappa}{2}+\mathrm{i}\omega_d\right)(p_1+p_2-\tau-r)}\\
&\ \ \ \times\int_{-\infty}^rd\tau_1\ e^{-\left( \frac{\kappa }{2}-\mathrm{i}\omega_d\right)(r-\tau_1)}\delta(\tau_1-\tau)\int_{-\infty}^rd\tau_{2}\\
&\ \ \ \times e^{-\left(\frac{\kappa }{2}+\mathrm{i}\omega_d \right) (r-\tau _{2})}[\xi _{2}(r)\xi _{1}(\tau_{2})+\xi _{1}(r)\xi _{2}(\tau _{2})].
\end{aligned}\end{equation}

It is hard to see that $\Gamma(p_{1},p_{2})$ in Eq. (\ref{eta_3}) is symmetric in the sense that $\Gamma(p_{1},p_{2})= \Gamma(p_{2},p_{1})$. In the following, we present a function  which is symmetric.  By Eq. (\ref{eq:b_out3}) we have
\begin{equation}\nonumber\begin{aligned}
&\left\langle 0g|b_{\mathrm{out}}(p_{1})b_{\mathrm{out}}(p_{2})b^{\dag
}(t_{1})b^{\dag }(t_{2})|0g\right\rangle\\
=&\; \frac{1}{2}\left\langle 0g|b_{\mathrm{out}}(p_{1})b_{\mathrm{out}}(p_{2})b^{\dag }(t_{1})b^{\dag }(t_{2})|0g\right\rangle
+\frac{1}{2}\left\langle 0g|b_{\mathrm{out}}(p_{2})b_{\mathrm{out}}(p_{1})b^{\dag}(t_{1})b^{\dag }(t_{2})|0g\right\rangle
\end{aligned}\end{equation}
Hence, we may rewrite Eq. \eqref{eq:feb23_temp1} as 
\begin{equation}\label{eq:june18_temp1}\begin{aligned}
\left\vert\Psi_{\mathrm{out}}\right\rangle
=&\frac{1}{2\sqrt{N_2}}\int_{-\infty}^{\infty}dp_1\int_{-\infty}^{\infty}dp_2\left\vert1_{p_1}1_{p_2}\right\rangle
\lim_{\begin{subarray}{c}t_0\rightarrow-\infty \\ t\rightarrow\infty\end{subarray}}\int_{t_0}^{t}dt_1\xi_1(t_1)\int_{t_0}^{t}dt_2\xi_2(t_2)\\
&\times\left\langle0g\right\vert b_{\mathrm{out}}(p_2)b_{\mathrm{out}}(p_1)b^{\dag}(t_1)b^{\dag}(t_2)\left\vert 0g\right\rangle.
\end{aligned}\end{equation}
Similar to the derivations for Eq. \eqref{eq:feb23_output} given above, Eq. \eqref{eq:june18_temp1} can be simplified as
\begin{equation}\label{eq:june18_output}
\left\vert\Psi _{\mathrm{out}}\right\rangle \\
= \frac{1}{2\sqrt{N_{2}}}\int_{-\infty }^{\infty }dp_{1}\int_{-\infty}^{\infty }dp_{2}\ \Gamma(p_2,p_1)  \left\vert 1_{p_{1}}1_{p_{2}}\right\rangle.
\end{equation}
Consequently, 
\begin{equation}\label{eq:june18b_output}
\left\vert\Psi _{\mathrm{out}}\right\rangle \\
= \frac{1}{2\sqrt{N_{2}}}\int_{-\infty }^{\infty }dp_{1}\int_{-\infty}^{\infty }dp_{2}\ \eta(p_1,p_2)  \left\vert 1_{p_{1}}1_{p_{2}}\right\rangle,
\end{equation}
where
\begin{equation}\label{eq:june18c}
 \eta(p_{1},p_{2}) = \frac{\Gamma(p_{1},p_{2})+\Gamma(p_{2},p_{1})}{2}.
\end{equation}
It is easy to see that $ \eta(p_{1},p_{2}) $ in Eq. \eqref{eq:june18c} is exactly that in Eq. \eqref{eta_5}.
The proof of Lemma \ref{lem:out4} is completed. $\square$

\begin{remark}
From Eq. (\ref{eta_5}), one can see that the output pulse shape contains $16$ terms. Interestingly, in the study of quantum filtering of a two-level system driven by the two-photon state $|2_{\xi_1,\xi_2}\rangle$, a system of $16$ ordinary differential equations are needed to represent the two-photon filter or the master equation \cite[Corollary 3.2]{SONG:13MULTI}. That is, there is consistency between output two-photon field state and two-photon quantum filtering.
\end{remark}

The expression of the steady-state output field state in Eq. (\ref{eta_5}) has $16$ terms, which looks rather complicated. In what follows, we further simplify $\eta (p_{1},p_{2})$ to get its form as given in Eq. \eqref{eta_may27}, thus completing the proof of Theorem \ref{thm:output}.

Firstly, it can be readily shown that
\footnotesize
\begin{equation}\label{temp_20a}\begin{aligned}
&\xi_1(p_2)\xi_2(p_1)+\xi_1(p_1)\xi_2(p_2)-\kappa\int_{-\infty}^{p_1}d\tau
e^{-\left(\frac{\kappa}{2}+\mathrm{i}\omega_d\right)(p_{1}-\tau )}[\xi_2(\tau)\xi_1(p_2)+\xi_1(\tau)\xi_2(p_2)]\\
=&g_G\ast\xi_1(p_1)\times\xi_2(p_2)+g_G\ast\xi_2(p_1)\times\xi_1(p_2).
\end{aligned}\end{equation}
\normalsize
Secondly,
\footnotesize
\begin{equation}\label{temp_20b}\begin{aligned}
&-\kappa\int_{-\infty}^{p_2}d\tau e^{-\left(\frac{\kappa}{2}+\mathrm{i}\omega_d\right)(p_2-\tau)}[\xi_1(p_1)\xi_2(\tau)+\xi_2(p_1)\xi_1(\tau)]\\
&+\kappa^2\int_{-\infty}^{p_2}d\tau\int_{-\infty}^{p_1}dre^{-\left(\frac{\kappa}{2}+\mathrm{i}\omega_d\right)(p_1+p_2-\tau-r)}
[\xi_1(r)\xi_2(\tau)+\xi_2(r)\xi_1(\tau)]\\
=&-\kappa\ g_G\ast\xi_1(p_1)\times\int_{-\infty}^{p_2}d\tau e^{-\left(\frac{\kappa}{2}+\mathrm{i}\omega_d\right)(p_2-\tau)}\xi_2(\tau)\\
&-\kappa\ g_G\ast\xi_2(p_{1})\times\int_{-\infty}^{p_2}d\tau e^{-\left(\frac{\kappa}{2}+\mathrm{i}\omega_d\right)(p_2-\tau)}\xi_1(\tau),
\end{aligned}\end{equation}
\normalsize
where Eq. (\ref{temp_20a}) is used in the last step.
%Because of
%\begin{equation}\begin{aligned}
%&g_G\ast\xi_1(p_1)\times\xi_2(p_2)-g_G\ast\xi_1(p_1)\times
%\kappa\int_{-\infty}^{p_2}d\tau e^{-\left(\frac{\kappa}{2}+\mathrm{i}\omega_d\right)(p_2-\tau)}\xi_2(\tau)\\
%=&g_G\ast\xi_1(p_1)\times g_G\ast\xi_2(p_2),
%\end{aligned}\end{equation}
%and
%\begin{equation}\begin{aligned}
%&g_G\ast\xi_2(p_1)\times\xi_1(p_2)-g_G\ast\xi_2(p_1)\times
%\kappa\int_{-\infty}^{p_2}d\tau e^{-\left(\frac{\kappa}{2}+i\omega_d\right)(p_2-\tau)}\xi_1(\tau)\\
%=&g_G\ast\xi_2(p_1)\times g_G\ast\xi_1(p_2).
%\end{aligned}\end{equation}
By adding (\ref{temp_20a}) and (\ref{temp_20b}), the first $8$ terms of $\eta(p_{1},p_{2})$ becomes
\begin{equation}\label{temp_20c}
g_G\ast\xi_1(p_1)\times g_G\ast\xi_2(p_2)+g_G\ast\xi_2(p_1)\times g_G\ast\xi_1(p_2).
\end{equation}

Thirdly, notice that the remaining $8$ terms of $\eta (p_{1},p_{2})$ (ignoring the common coefficient $\kappa^2$) can be  simplified to
\footnotesize
\begin{equation}\label{temp_21c_v2}\begin{aligned}
&\int_{-\infty}^{p_1}d\tau e^{-\left(\frac{\kappa}{2}+\mathrm{i}\omega_d\right)(p_1-\tau)}\int_{-\infty}^{\tau}d\tau_1e^{-\left(\frac{\kappa}{2}-\mathrm{i}\omega_d\right)(\tau-\tau_1)}
g_G\ast\delta(p_2-\tau_1)\\
&\times\left[\xi_1(\tau)\int_{-\infty}^{\tau}d\tau_2e^{-\left(\frac{\kappa}{2}+\mathrm{i}\omega_d\right)(\tau-\tau_2)}\xi_2(\tau_2)\right. \left.+\xi_2(\tau)\int_{-\infty}^{\tau}d\tau_2e^{-\left(\frac{\kappa}{2}+\mathrm{i}\omega_d\right)(\tau-\tau_2)}\xi_1(\tau_2)\right] \\
+&\int_{-\infty}^{p_2}d\tau e^{-\left(\frac{\kappa}{2}+\mathrm{i}\omega_d\right)(p_2-\tau)}\int_{-\infty}^{\tau}d\tau_1e^{-\left(\frac{\kappa}{2}-\mathrm{i}\omega_d\right)(\tau-\tau_1)}
g_G\ast\delta(p_1-\tau_1)\\
&\times\left[\xi_1(\tau)\int_{-\infty}^{\tau}d\tau_2e^{-\left(\frac{\kappa}{2}+\mathrm{i}\omega_d\right)(\tau-\tau_2)}\xi_2(\tau_2)\right. \left.+\xi_2(\tau)\int_{-\infty}^{\tau}d\tau_2e^{-\left(\frac{\kappa}{2}+\mathrm{i}\omega_d\right)(\tau-\tau_2)}\xi_1(\tau_2)\right],
\end{aligned}\end{equation}
\normalsize
where the fact
\begin{equation}\label{key_9a}\begin{aligned}
\delta(p_2-\tau_1)-\kappa\int_{-\infty}^{p_2}ds\ e^{-\left(\frac{\kappa}{2}+i\omega_d\right)(p_2-s)}\delta(\tau_1-s)
=g_G\ast\delta(p_2-\tau_1)
\end{aligned}\end{equation}
is used in the derivation. Moreover, the first term in Eq. (\ref{temp_21c_v2})  can be simplified as
\footnotesize
\begin{equation}\label{May28}\begin{aligned}
&\kappa^2\int_{-\infty}^{p_1}d\tau e^{-\left(\frac{\kappa}{2}+\mathrm{i}\omega_d\right)(p_1-\tau)}\int_{-\infty}^{\tau}d\tau_1e^{-\left(\frac{\kappa}{2}-\mathrm{i}\omega_d\right)(\tau-\tau_1)}
g_G\ast\delta(p_2-\tau_1) \\
&\times\left[\xi_1(\tau)\int_{-\infty}^{\tau}d\tau_2e^{-\left(\frac{\kappa}{2}+\mathrm{i}\omega_d\right)(\tau-\tau_2)}\xi_2(\tau_2)\right. \left.+\xi_2(\tau)\int_{-\infty}^{\tau}d\tau_2e^{-\left(\frac{\kappa}{2}+\mathrm{i}\omega_d\right)(\tau-\tau_2)}\xi_1(\tau_2)\right] \\
%=&\kappa\int_{-\infty}^{p_1}d\tau e^{-\left(\frac{\kappa}{2}+\mathrm{i}\omega_d\right)(p_1-\tau)}\int_{-\infty}^{\tau}d\tau_1e^{-\left(\frac{\kappa}{2}-\mathrm{i}\omega_d\right)(\tau-\tau_1)}
%g_G\ast\delta(p_2-\tau_1) \\
%&\times\left[\xi_1(\tau)\kappa\int_{-\infty}^{\tau}d\tau_2e^{-\left(\frac{\kappa}{2}+\mathrm{i}\omega_d\right)(\tau-\tau_2)}\xi_2(\tau_2)\right. +\xi_2(\tau)\kappa\int_{-\infty}^{\tau}d\tau_2e^{-\left(\frac{\kappa}{2}+\mathrm{i}\omega_d\right)(\tau-\tau_2)}\xi_1(\tau_2) \\
%&+2\xi_1(\tau)\xi_2(\tau)-2\xi_1(\tau)\xi_2(\tau)\bigg] \\
=&\bigg\{\begin{array}{cc} 2\kappa\ e^{-\frac{\kappa}{2}(p_1-p_2)-\mathrm{i}\omega_d(p_1+p_2)}\int_{p_2}^{p_1}d\tau\ e^{2\mathrm{i}\omega_d\tau}\left[\xi_1(\tau)\xi_2(\tau)-\frac{\xi_1(\tau)\nu_2(\tau)+\nu_1(\tau)\xi_2(\tau)}{2}\right], &p_1\geq p_2,\\
0, & p_1<p_2.
\end{array}
\end{aligned}\end{equation}
\normalsize
The second term in  Eq. (\ref{temp_21c_v2}) can be treated in a similar way. Finally, by Eq. (\ref{temp_20c}), Eq. (\ref{temp_21c_v2}) and Eq. (\ref{May28}), we get $\eta(p_1,p_2)$ as given in Eq. \eqref{eta_may27}.  The proof of Theorem \ref{thm:output} is completed. $\square$

\section{The derivation of Eqs. (\ref{2_key_a})-(\ref{2_key_d})}\label{appendixd}

By Eq. \eqref{sys_3a}, we have
\begin{equation}\label{mar14_1a}\begin{aligned}
\left\langle0g\right\vert\sigma_-(r)b_k^{\dagger}(q)\left\vert0g\right\rangle
=-\sqrt{\kappa_k}\left[\delta_{1k}+\delta_{2k}\right]\int_{t_0}^rd\tau\ e^{-\frac{\kappa_1+\kappa_2}{2}(r-\tau)}\delta(\tau-q).%k=1,2,
\end{aligned}\end{equation}

By Eq. (\ref{mar14_1a}), we can show that
\begin{equation}\label{mar16_2}\begin{aligned}
&\left\langle0g\right\vert b_j(l)\sigma_z(r)b_k^{\dagger}(t)\left\vert0g\right\rangle\\
=&2\sqrt{\kappa_j\kappa_k}\int_{t_0}^rd\tau_1\ e^{-\frac{\kappa_1+\kappa_2}{2}(r-\tau_1)}\delta(\tau_1-l)
\int_{t_0}^rd\tau_2\ e^{-\frac{\kappa_1+\kappa_2}{2}(r-\tau_2)}\delta(\tau_2-t)\\
&-\delta_{jk}\delta(l-t),\ \ \ j,k=1,2.
\end{aligned}\end{equation}
Then, we can conclude the following result.
\footnotesize
\begin{equation*}\begin{aligned}
&\left\langle0g\right\vert b_1(l)\sigma_z(r)b_1(r)b_1^{\dagger}(t_1)b_2^{\dagger}(t_2)\left\vert0g\right\rangle\\
=&\; 2\sqrt{\kappa_1\kappa_2}\delta(r-t_1)\int_{t_0}^rd\tau_1\ e^{-\frac{\kappa_1+\kappa_2}{2}(r-\tau_1)}\delta(\tau_1-l)
\int_{t_0}^rd\tau_2\ e^{-\frac{\kappa_1+\kappa_2}{2}(r-\tau_2)}\delta(\tau_2-t_2),\\
&\left\langle0g\right\vert b_2(l)\sigma_z(r)b_2(r)b_1^{\dagger}(t_1)b_2^{\dagger}(t_2)\left\vert0g\right\rangle\\
=&\; 2\sqrt{\kappa_1\kappa_2}\delta(r-t_2)\int_{t_0}^rd\tau_1\ e^{-\frac{\kappa_1+\kappa_2}{2}(r-\tau_1)}\delta(\tau_1-l)
\int_{t_0}^rd\tau_2\ e^{-\frac{\kappa_1+\kappa_2}{2}(r-\tau_2)}\delta(\tau_2-t_1),\\
&\left\langle0g\right\vert b_1(l)\sigma_z(r)b_2(r)b_1^{\dagger}(t_1)b_2^{\dagger}(t_2)\left\vert0g\right\rangle\\
=&\; 2\sqrt{\kappa_1\kappa_1}\delta(r-t_2)\int_{t_0}^rd\tau_1\ e^{-\frac{\kappa_1+\kappa_2}{2}(r-\tau_1)}\delta(\tau_1-l)
\int_{t_0}^rd\tau_2\ e^{-\frac{\kappa_1+\kappa_2}{2}(r-\tau_2)}\delta(\tau_2-t_1)\\
&-\delta(l-t_1)\delta(r-t_2),\\
&\left\langle0g\right\vert b_2(l)\sigma_z(r)b_1(r)b_1^{\dagger}(t_1)b_2^{\dagger}(t_2)\left\vert0g\right\rangle\\
=&\; 2\sqrt{\kappa_2\kappa_2}\delta(r-t_1)\int_{t_0}^rd\tau_1\ e^{-\frac{\kappa_1+\kappa_2}{2}(r-\tau_1)}\delta(\tau_1-l)
\int_{t_0}^rd\tau_2\ e^{-\frac{\kappa_1+\kappa_2}{2}(r-\tau_2)}\delta(\tau_2-t_2)\\
&-\delta(l-t_2)\delta(r-t_1).
\end{aligned}\end{equation*}
\normalsize
Consequently, we have the following expressions of the quantities in Eqs. (\ref{2_key_a})-(\ref{2_key_d}).
\footnotesize
\begin{equation}\label{eq:may27_10}\begin{aligned}
&\langle0g|b_{\mathrm{out},i}(p_1)b_{\mathrm{out},j}(p_2)b_1^\dagger(t_1)b_2^\dagger(t_2)|0g\rangle\\
=&\; \kappa_j\int_{-\infty}^{p_2}dr\ e^{-\frac{\kappa_1+\kappa_2}{2}(p_2-r)}
\langle0g|g_{G_{ii}}\ast b_i(p_1)\sigma_z(r)b_j(r)b_1^\dagger(t_1)b_2^\dagger(t_2)|0g\rangle\\
&+\sqrt{\kappa_1\kappa_2}\int_{-\infty}^{p_2}dr\ e^{-\frac{\kappa_1+\kappa_2}{2}(p_2-r)}
\langle0g|g_{G_{ii}}\ast b_i(p_1)\sigma_z(r)b_{\frac{2}{j}}(r)b_1^\dagger(t_1)b_2^\dagger(t_2)|0g\rangle\\
&-\kappa_j\sqrt{\kappa_1\kappa_2}\int_{-\infty}^{p_2}dr\ e^{-\frac{\kappa_1+\kappa_2}{2}(p_2-r)}\int_{-\infty}^{p_1}d\tau\ e^{-\frac{\kappa_1+\kappa_2}{2}(p_1-\tau)}\\
&\ \ \ \times\langle0g|b_{\frac{2}{i}}(\tau)\sigma_z(r)b_j(r)b_1^\dagger(t_1)b_2^\dagger(t_2)|0g\rangle\\
&-\kappa_1\kappa_2\int_{-\infty}^{p_2}dr\ e^{-\frac{\kappa_1+\kappa_2}{2}(p_2-r)}\int_{-\infty}^{p_1}d\tau\ e^{-\frac{\kappa_1+\kappa_2}{2}(p_1-\tau)}\\
&\ \ \ \times\langle0g|b_{\frac{2}{i}}(\tau)\sigma_z(r)b_{\frac{2}{j}}(r)b_1^\dagger(t_1)b_2^\dagger(t_2)|0g\rangle\\
&-\sqrt{\kappa_i\kappa_{\frac{2}{j}}}\delta(p_2-t_j)\int_{-\infty}^{p_1}d\tau\ e^{-\frac{\kappa_1+\kappa_2}{2}(p_1-\tau)}\delta(\tau-t_{\frac{2}{j}})\\
&+(1-\delta_{ij})\delta(p_2-t_j)\delta(p_1-t_i), ~~ i,j=1,2.
\end{aligned}\end{equation}
\normalsize
Substituting Eq. \eqref{eq:may27_10} into Eq. \eqref{mar17_1}  gives Eq. \eqref{June4}. $\square$

%
%\section*{Acknowledgments}\label{Ack}
%The authors would like to thank Hongchen Fu, Zhanfeng Jiang, Jun Jing and Bobo Wei for their very helpful discussions.

%%\bibliographystyle{siamplain}
%\bibliography{references}

\bibliographystyle{plain}        % Include this if you use bibtex
\bibliography{references}

\end{document}